\documentclass[twocolumn]{aastex63}

\usepackage{natbib}
\usepackage{graphicx}

\usepackage[caption=false]{subfig}
\let\tablenum\relax

\usepackage[version-1-compatibility]{siunitx}
\DeclareSIUnit\parsec{pc}
\usepackage{soul}
\received{}
\revised{}
\accepted{\today}
\submitjournal{AJ}
\usepackage{lineno}
\usepackage{lipsum}
\usepackage{booktabs}
\usepackage{textgreek}
\usepackage{threeparttable}

\shorttitle{HESS J1640$-$465 and HESS J1641$-$463}
\shortauthors{A.-R.Mar\`es et al.}

\DeclareUnicodeCharacter{2212}{-}
\begin{document}

\title{Constraining the origin of the puzzling source HESS J1640$-$465 and the PeVatron candidate HESS J1641$-$463 using \textit{Fermi}-LAT observations}

\author{A.~Mares}
\email{mares@cenbg.in2p3.fr}
\affiliation{Univ. Bordeaux, CNRS, CENBG, UMR 5797, F-33170 Gradignan, France}
\author{M.~Lemoine-Goumard}
\email{lemoine@cenbg.in2p3.fr}
\affiliation{Univ. Bordeaux, CNRS, CENBG, UMR 5797, F-33170 Gradignan, France}
\author{F.~Acero}
\affiliation{AIM, CEA, CNRS, Universit\'e Paris-Saclay, Universit\'e Paris Diderot, Sorbonne Paris Cit\'e, F-91191 Gif-sur-Yvette, France}
\author{C.~J.~Clark}
\affiliation{Jodrell Bank Centre for Astrophysics, School of Physics and Astronomy, The University of Manchester, M13 9PL, UK}
\author{J.~Devin}
\affiliation{Univ. Bordeaux, CNRS, CENBG, UMR 5797, F-33170 Gradignan, France}
\author{S.~Gabici}
\affiliation{AIM, CEA, CNRS, Universit\'e Paris-Saclay, Universit\'e Paris Diderot, Sorbonne Paris Cit\'e, F-91191 Gif-sur-Yvette, France}
\author{J.~D.~Gelfand}
\affiliation{New York University Abu Dhabi, Abu Dhabi, United Arab Emirates}
\author{D.~A.~Green}
\affiliation{Cavendish Laboratory, Cambridge CB3 0HE, UK}
\author{M.-H.~Grondin}
\affiliation{Univ. Bordeaux, CNRS, CENBG, UMR 5797, F-33170 Gradignan, France}

\begin{abstract}
This list is preliminary; the status is not yet "ready to submit"
\end{abstract}



\begin{abstract}

There are only few very-high-energy sources in our Galaxy which might accelerate particles up to the knee of the cosmic-ray spectrum. To understand the mechanisms of particle acceleration in these PeVatron candidates, \textit{Fermi}-LAT  and H.E.S.S. observations are essential to characterize their $\gamma$-ray emission. HESS J1640$-$465 and the PeVatron candidate HESS J1641$-$463 are two neighboring (\ang[astroang]{0.25}) $\gamma$-ray sources, spatially coincident with the radio supernova remnants (SNRs) G338.3$-$0.0 and G338.5+0.1. Detected both by H.E.S.S. and \textit{Fermi}-LAT, we present here a morphological and spectral analysis  of these two sources using 8 years of \textit{Fermi}-LAT data between 200 \si{\mega\electronvolt} and 1 \si{\tera\electronvolt}   with multi-wavelength observations to assess their nature. 
The morphology of HESS J1640$-$465 is described by a 2D Gaussian ($\sigma=$ \ang[astroang]{0.053} $\pm$  \ang[astroang]{0.011}$_{stat}$ $ \pm$ \ang[astroang]{0.03}$_{syst}$) and its spectrum is modeled by a power-law with a spectral index $\Gamma = 1.8\pm0.1_{\rm stat}\pm0.2_{\rm syst}$.
HESS J1641$-$463 is detected as a point-like source and its GeV emission is described by a logarithmic-parabola spectrum with $\alpha = 2.7 \pm 0.1_ {\rm stat} \pm 0.2_ {\rm syst} $ and significant curvature of $\beta = 0.11 \pm 0.03_ {\rm stat} \pm 0.05_ {\rm syst} $. Radio and X-ray flux upper limits were derived. We investigated scenarios to explain their emission, namely the emission from accelerated particles within the SNRs spatially coincident with each source,  molecular clouds illuminated by cosmic rays from the close-by SNRs, and a pulsar/PWN origin. Our new \emph{Fermi}-LAT results and the radio and flux X-ray upper limits pose severe constraints on some of these models.
%For HESS J1641$-$463, the GeV emission could be either an hadronic emission from the coincident SNR G338.5+0.1 or a pulsar emission. In this latter case, the TeV emission can be explained by the PWN powered by this putative pulsar. We also investigated the case of escaping protons from the SNR G338.3$-$0.0 (spatially coincident with HESS J1640$-$465) interacting with molecular cloud to explain the TeV emission of HESS J1641$-$463. For both sources, we studied the broad-band spectral energy distribution from radio to $\gamma$-rays of the region in order to provide new constraints concerning the identification of these two $\gamma$-ray sources.  Assuming a distance of 11 kpc, the emission of both sources cannot be easily explain by one simple mechanism. The multi-wavelength emission of both sources combined with multiples physical model that we have test does not allow us to favor one model rather than another.

\end{abstract}


\vspace{4.1cm}

\section{Introduction} 
\label{sec:intro}

HESS J1640$-$465 is a very-high-energy (VHE) $\gamma$-ray source located in the Galaxy discovered by the High-Energy Stereoscopic System (H.E.S.S.) during its Galactic Plane Survey \citep{aharonian_h.e.s.s._2006}. First detected as a point source, further observations revealed a Gaussian extension of $\sigma =4.32\arcmin \pm 0.18 \arcmin $, with a spectrum described by an exponential cut-off power-law \citep[$\Gamma=2.11 \pm 0.09$;][]{abramowski_hess_2014} with a cut-off energy: $E_{\rm c}= 6.0\pm^{2.0}_{−1.2}$ TeV. HESS J1640$-$465 coincides with a dense molecular cloud (G338.4$+$0.1) overlapping the north western part of the supernova remnant (SNR) shell of G338.3$-$0.0 \citep{abramowski_hess_2014}. The age of this radio shell-type SNR \citep[$\sigma =7'$;][]{whiteoak_most_1996} remains uncertain due to the large amount of surrounding interstellar material that could impact the evolution of the SNR, and is estimated to be between 1 and 8 \si{\kilo}yr. X-ray observations of the region with \textit{XMM-Newton} \citep{funk_xmm-newton_2007} revealed an extended non-thermal source ($\sigma=\ang{;;27} \pm \ang{;;3}$)  close to the center of the remnant and coincident with the \si{\tera\electronvolt} emission. Further observations with \textit{Chandra} \citep{lemiere_high-resolution_2009} found a compact X-ray object which was suggested to be the associated pulsar. No X-ray emission was detected from the SNR G338.3$-$0.0 because of the low temperature of the SNR ejecta and/or the large column density in the line of sight. Later, X-ray pulsations were detected by \cite{gotthelf_nustar_2014} at the position of this compact object using the \textit{Nuclear Spectroscopic Telescope Array (NuSTAR)}. Later on, \cite{archibald16} calculated the braking index of this pulsar using a 2.3 year long phase-connected timing solution. With a characteristic age of $\tau_{c}=3350$ years and a spin-down power of $\dot{E}$ = $4.4\times10^{36}$ erg \si{\per\second}, the pulsar PSR J1640$-$4631 has similar properties to pulsars powering \si{\tera\electronvolt} pulsar wind nebulae \citep[PWNe;][]{klepser_population_2013}.  No
radio counterpart was found, and only upper limits on the radio fluxes in this region were derived.
 No $\gamma$-ray or radio pulsation has been detected so far.

A spatially coincident \si{\giga\electronvolt} $\gamma$-ray source (1FGL J1640.8$-$465) was then revealed by the \textit{Fermi} Large Area Telescope (LAT) with a soft spectral index $\Gamma = 2.3\pm0.1$ \citep{slane_fermi_2010}.  A distance of 8 -- 13 \si{\kilo\parsec} was derived based on 21 \si{\centi\meter} HI absorption by  \cite{lemiere_high-resolution_2009}, which implied that HESS J1640$-$465 was the most luminous VHE $\gamma$-ray  source  known  in  the  Galaxy at that time. The later discovery of the neighboring VHE source HESS J1641$-$463 \citep{abramowski_discovery_2014} changed the \si{\giga\electronvolt} soft spectral index from 2.3  to a harder spectral index of $\Gamma$= 1.99 $\pm$ 0.04 \citep{lemoine-goumard_hess_2014} from a dedicated analysis of both sources in the 0.1 -- 300 GeV energy band.   Using high-resolution multi-frequency radio data of G338.3$-$0.0, \cite{castelletti_first_2011} derived upper limits on the radio flux from a potential extended radio nebula as the multi-frequency radio
observations did not detect radio pulsations toward the point
source. More recently,  \cite{xin_hess_2018} obtained a hard index of $\Gamma=1.42\pm0.19$ during a dedicated analysis of HESS J1640$-$465 (from 10 to 500 \si{\giga\electronvolt}).

 In addition to HESS J1640$-$465, there is a nearby (\ang[astroang]{0.25}) weak source HESS J1641$-$463, which could only be detected by H.E.S.S. using an energy threshold of 4 \si{\tera\electronvolt} \citep{abramowski_discovery_2014}. Characterized as a point-like source with a spectral index of $\Gamma$ = 2.07 $\pm$ 0.11 and significant detection of photons up to 30 \si{\tera\electronvolt}, HESS J1641$-$463 is one of the best H.E.S.S. PeVatron candidates. Its location is coincident with the radio SNR G338.5$+$0.1 whose age is estimated between 1.1 and 3.5 \si{\kilo}yr or between 5 and 17 \si{\kilo}yr \citep{Kothes2007}. The 843 \si{\mega\hertz} radio data from MOST revealed a bright dense HII region (G338.4$+$0.0) that connects in projection the two SNRs G338.3$-$0.0 and G338.5$+$0.1 \citep{whiteoak_most_1996}. No other known counterpart is found to be compatible with HESS J1641$-$463. \cite{lemoine-goumard_hess_2014} reported the discovery of a \si{\giga\electronvolt} source using 5 years of the \textit{Fermi}-LAT data  coinciding with this \si{\tera\electronvolt} source. This analysis shows a spectral index of $\Gamma = 2.47 \pm 0.05 $ for HESS J1641$-$463 in the 0.1 -- 300 \si{\giga\electronvolt} energy range, which differs significantly from the spectral shape in the TeV energy range ($\Gamma \sim 2.1$).
 
 At first, the detected \si{\tera\electronvolt} emission of HESS J1640$-$465 was interpreted as accelerated particles in the PWN  emitting $\gamma$-rays through inverse Compton scattering. This scenario was reinforced by the detection of the extended X-ray source and the associated X-ray pulsar PSR J1640$-$4631.
 \cite{xin_hess_2018} showed that the leptonic hypothesis of particle
acceleration in a PWN was the best model to reproduce the $\gamma$-ray emission of HESS J1640$-$465. They modeled the particles with a
power-law in the high-energy (HE) part of their spectrum. The cutoff in the \si{\tera\electronvolt} energy band seen by H.E.S.S. was then easily explained by the Klein-Nishima regime.

 Then, taking into account the radio data, the \si{\tera\electronvolt} emission of both neighboring sources could be explained by a self-consistent scenario in which protons might be accelerated  up to hundreds of \si{\tera\electronvolt} in either G338.5$+$0.1 or G338.3$-$0.0.  
 In this case, the observed broadband spectrum of HESS J1640$-$465  would come from  proton-proton interactions inside the dense HII region G338.4$+$0.1 located in the northern part of the SNR shell coincident with HESS J1640$-$465, as first proposed by \cite{abramowski_discovery_2014} and then studied with more  details by \cite{tang_self-consistent_2015}. Protons of higher energy could then escape from the accelerating zone in the SNR shock wave, travel toward the molecular cloud coinciding with HESS J1641$-$463 and interact with it. This would explain the relatively high brightness of the $\gamma$-ray emission from HESS J1641$-$463 in comparison with HESS J1640$-$465 above 4 \si{\tera\electronvolt}.  This scenario is possible if both sources are at the same distance. This hypothesis is supported by the radio ``bridge'', detected by \cite{whiteoak_most_1996}, connecting  HESS J1640$-$465 and HESS J1641$-$463. 
 
 %Nevertheless, the leptonic scenario cannot be excluded. Indeed, the TeV $\gamma$-emission of HESS J1641$-$463 might still be produced by accelerated protons interacting with the interstellar medium, while the GeV detection could come from a contaminating pulsar in the line of sight. 
 
 Another possibility is that HESS J1641$-$463 could be a binary system, with a pulsar-like spectrum at \si{\giga\electronvolt} energies that does not connect to the \si{\tera\electronvolt} spectrum coming from the wind shocks. However, no $\gamma$-ray variability has been detected using the \textit{Fermi}-LAT data \citep{gotthelf_nustar_2014}, and no star is seen at any wavelength with a position close to HESS J1641$-$463. \\
 
This work presents an analysis of HESS J1640$-$465 and HESS J1641$-$463 using 8 years of \textit{Fermi}-LAT data. Section \ref{sec:data} describes the \textit{Fermi}-LAT data used for the analysis, while Sections \ref{sec:morp} and \ref{sec:spec} present the results obtained from a morphological and spectral analysis of HESS J1640$-$465 and HESS J1641$-$463. Finally, in Section \ref{sec:multi}, we discuss the main implications of these results concerning the origin of the detected $\gamma$-ray signal. Our extensive analyses provide new constraints on the origin of the $\gamma$-ray emission as well as the efficiency of these two \si{\tera\electronvolt} sources to accelerate protons and contribute to the Galactic cosmic-ray flux around the knee.

\section{\textit{Fermi}-LAT analysis of HESS J1640\texorpdfstring{$-$}{-}465 and HESS J1641\texorpdfstring{$-$}{-}463}

\subsection{\textit{Fermi}-LAT observation and data analysis}
\label{sec:data}
The \textit{Fermi} Large Area Telescope (LAT) is an electron/positron pair conversion telescope launched in 2008 and observing the $\gamma$-ray sky between 20 \si{\mega\electronvolt} to more than 500 \si{\giga\electronvolt} \citep{atwood_large_2009}. The following analyses take full advantage of the data quality improvement provided by the Pass 8 data upgrade (\citealt{atwood_pass_2013, bruel_fermi-lat_2018}). Using 8 years of data, we performed morphological and spectral analyses above 1 \si{\giga\electronvolt} and 200 \si{\mega\electronvolt} respectively, and up to 1 \si{\tera\electronvolt}  for both sources within a  \ang[astroang]{15}$\times$\ang[astroang]{15}  region centered on HESS J1640$-$465. For these studies, we have used all $\gamma$-rays, except those coming from a zenith angle above  \ang[astroang]{90}  to the detector axis to avoid Earth's limb contamination \citep{abdo_fermi_2009}. Time intervals when \textit{Fermi} overflew the South Atlantic Anomaly were also removed. 
Version \textit{11-07-00} of the ScienceTools was used, together with the \textit{Fermipy} \citep{wood_Fermipy:_2017} package version \textit{0.18.0}. We used the version \textit{P8R3\_SOURCE\_V2} of the IRFs, released by the \textit{Fermi}-LAT collaboration, with the SOURCE event-class. This event-class is the most favorable for analysis of extended and point-like sources on medium to long time scales, while providing lower charged particle residual background contamination. 
Both analyses were made using \textit{gtlike}. Included in the \textit{Fermi} ScienceTools\footnote{The documentation of the Fermi ScienceTools is available at \textit{{\scriptsize https://fermi.gsfc.nasa.gov/ssc/data/analysis/documentation/}}}, \textit{gtlike} performs unbinned and binned maximum likelihood analyses \citep{mattox_likelihood_1996} by fitting a source model to the \textit{Fermi}-LAT data. The diffuse emission in the Galaxy is described by the standard LAT diffuse model \textit{gll\_iem\_v07.fits} calculated using templates from GALPROP\footnote{GALPROP is a numerical code for calculating the propagation of relativistic charged particles and the diffuse emissions produced during their propagation. The GALPROP code incorporates as much realistic astrophysical input as possible together with latest theoretical developments. The code calculates the propagation of cosmic-ray nuclei, antiprotons, electrons and positrons, and computes diffuse $\gamma$-rays and synchrotron emission in the same framework.} \citep{moskalenko_galprop_2017}. Extragalactic diffuse $\gamma$-rays, unresolved extragalactic sources and residual cosmic-ray emission are modeled by the isotropic spectral template in the analysis (\textit{iso\_P8R3\_SOURCE\_V2\_v1.txt}).\footnote{Available from the Fermi Science Support Center: \textit{{\scriptsize https://fermi.gsfc.nasa.gov/ssc/data/access/lat/BackgroundModels}}} During this analysis, the normalisation of the diffuse Galactic and isotropic emission were fitted.
Our maximum likelihood analysis combines the four PSF event-types\footnote{Each PSF event types indicate the quality of the reconstructed direction, the data is divided into quartiles, from the lowest quality quartile (PSF0) to the best quality quartile (PSF3).} in a joint likelihood function. Using the catalog of the \textit{Fermi}-LAT collaboration \citep[hereafter referred to as the 4FGL;][]{the_fermi-lat_collaboration_fermi_2019} based on more than 8 years of \textit{Fermi}-LAT data, we took into account sources up to  \ang[astroang]{20} from HESS J1640$-$465. The spectral parameters of sources within  \ang[astroang]{3.5} of the center are left free, while the others are fixed at the value of the 4FGL catalog. Energy dispersion was taken into account for both morphological and spectral analyses. \\

\subsection{Morphological analysis}
\label{sec:morp}

As the Point Spread Function (PSF) of the LAT improves with energy, we chose to perform the morphological analysis of HESS J1640$-$465 and HESS J1641$-$463 above 1 \si{\giga\electronvolt} to benefit from  good angular resolution, without loosing too many photons.

\begin{figure}[!tbp]
  \centering
  \includegraphics[scale=0.04]{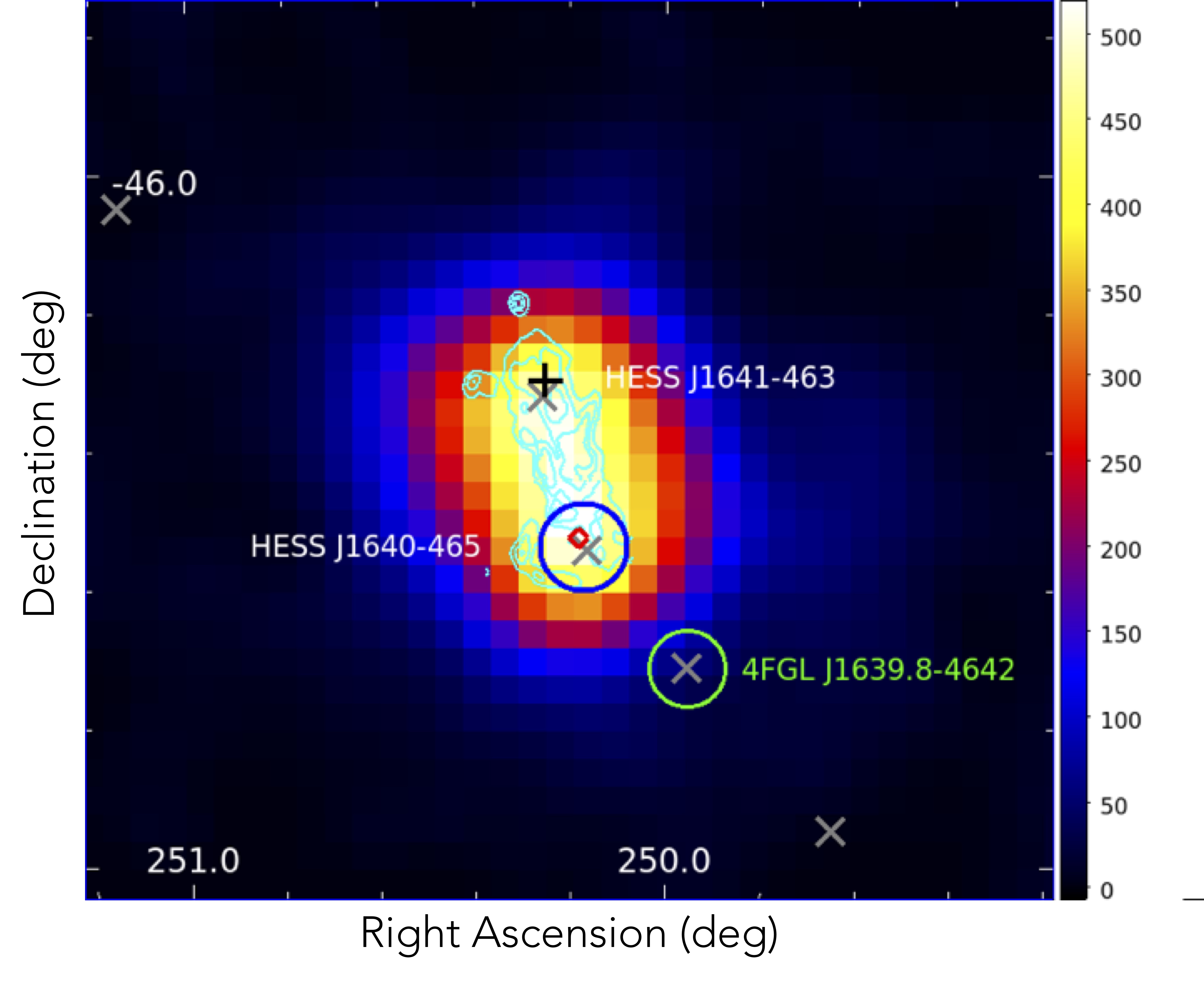}
  \hspace{0.cm}
  \includegraphics[scale=0.04]{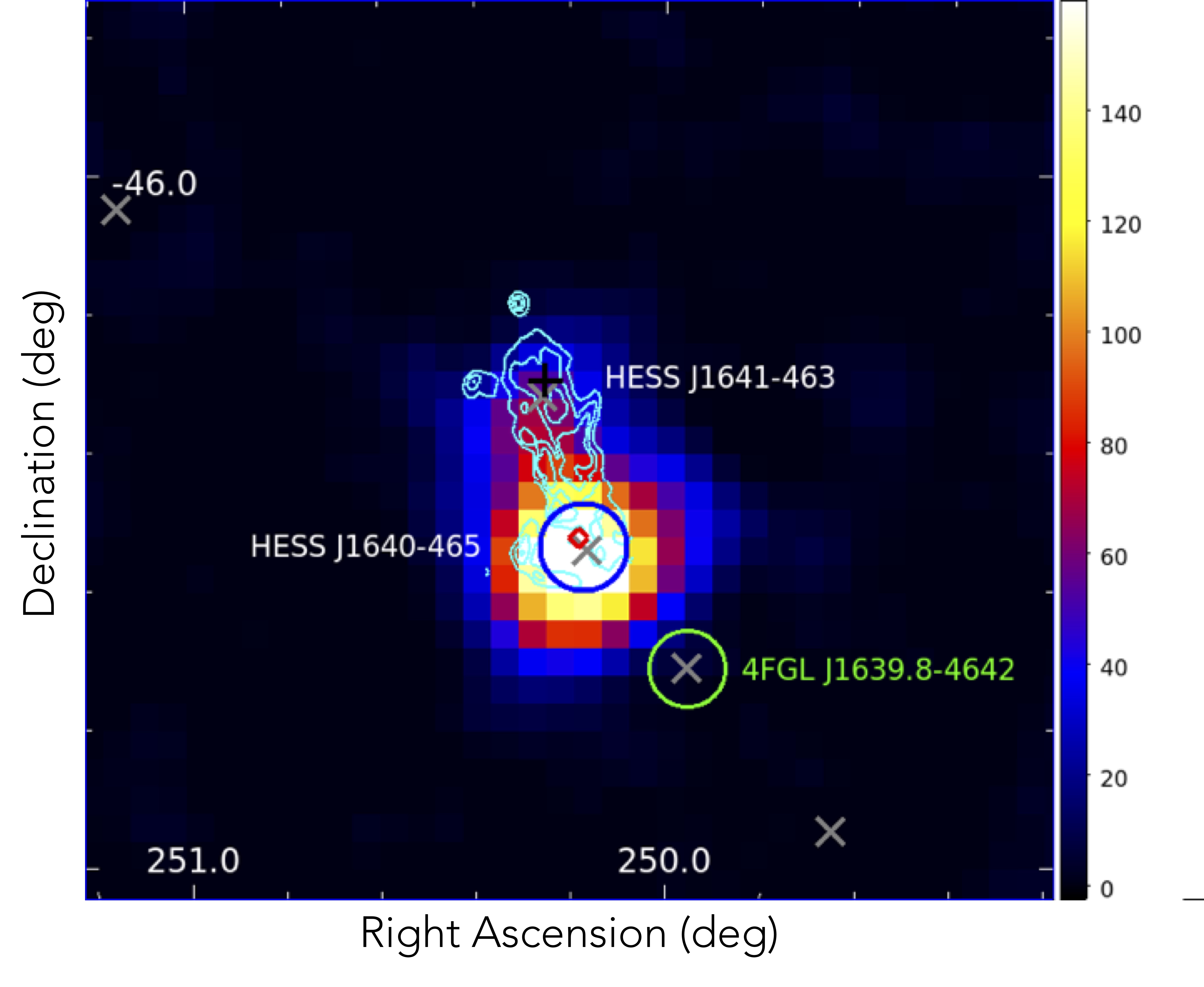}
  \caption{\textit{Fermi}-LAT residual TS maps of a \ang[astroang]{1.5} $\times$  \ang[astroang]{1.5} region around HESS J1640$-$465 above 1 \si{\giga\electronvolt} (Top) and above 10 \si{\giga\electronvolt} (Bottom). Sources from the 4FGL catalog are indicated with grey crosses ($\times$). Our best localisation of the point source HESS J1641$-$463 is marked with the black plus ($+$), while the best localisation and extension of the Gaussian HESS J1640$-$465 are represented by the blue circle. The red diamond indicates the position of the pulsar PSR J1640$-$4631. Radio continuum at 843 \si{\mega\hertz} is shown as cyan contours \citep{whiteoak_most_1996}. All sources are included in the background model except HESS J1640$-$465 and HESS J1641$-$463. The green circle indicates the position of the closest 4FGL source 4FGL J1639.8$-$4642 discussed in Section \ref{sec:spec}.}
\label{fig:ts}
\end{figure}

Two sources are detected in the 4FGL catalog (4FGL J1640.6$-$4632 and 4FGL J1641.0$-$4619), whose position coincides with HESS J1640$-$465 and HESS J1641$-$463. Figure \ref{fig:ts} shows images of the sources in two energy bands (  top: E$>$1GeV; bottom: E$>$10GeV) showing that HESS J1641$-$463 only appears below 10~GeV. Using \textit{Fermipy}, we have computed the associated residual Test Statistic map of the region, with a binning of  \ang[astroang]{0.04}. The Test Statistic (TS) is defined as twice the log-likelihood difference between the model with a putative additional source and the reference model. It follows a chi-square distribution with N (number of parameters associated to the additional source) degrees of freedom assuming the reference model and the N parameters are non-degenerate. This TS map was obtained by testing, at the center of each pixel, the presence of a point-like source with a power-law spectrum with a fixed index of 2. The map contains the TS value for each bin and gives the significance of the detection in excess of the background at a given position. We then looked for potential new sources in the region by fitting the prefactors and the indices of the hotspots  with TS above 16. If the TS of a source was above 25 after the fit, we added them in the final source model of the region. Six sources were found by this method above 1 \si{\giga\electronvolt}.  Two additional sources were added this way above 200 \si{\mega\electronvolt} (see Table \ref{table:add}). One of the added sources, PS J1703.2$-$4145, is close to a source (4FGL J1704.1$-$4140) detected in the second release of the 4FGL catalog \citep{ajello_fourth_2020}. It should be noted here that, unlike our analysis, the 4FGL introduced weights that give more importance to high energy photons compared to the low energy ones. This allows to mitigate the effect of systematic errors due to our imperfect knowledge of the Galactic diffuse emission but decreases the TS value of sources with a soft spectrum. Since the sources reported in Table \ref{table:add} have relatively soft spectra and TS below 98, they could have been missed due to the use of weights in the 4FGL and 4FGL-DR2 catalogs. It might also be that these sources are artificial structures from the diffuse emission model in this region. A more careful analysis would be needed to confirm these soft and faint sources.

\begin{table*}
\centering
\caption{Best morphological parameters (positions and extension) for HESS J1640$-$465 and HESS J1641$-$463.}
\label{table:best_morpho}
\begin{tabular}{ccccc}
\multicolumn{5}{c}{} \\
\hline
& & \textit{FermiPy} position (A)  & \textit{Fermipy} best fits (B) & H.E.S.S. position/extension (C) \\
HESS J1640$-$465 & R.A. ($^{\circ}$) & $250.18 \pm 0.01_{\rm stat}$ &  $250.17 \pm 0.01_{\rm stat} \pm 0.01_{\rm syst}$ & $250.180$ \\
& Dec ($^{\circ}$) &  $-46.55 \pm 0.01_{\rm stat}$ & $-46.55 \pm 0.01_{\rm stat} \pm 0.01_{\rm syst}$ & $-46.530$ \\
& $\sigma$ ($^{\circ}$) & $-$ & $0.053 \pm 0.011_{\rm stat} \pm 0.008_{\rm syst}$ & $0.072$   \\
HESS J1641$-$463 & R.A. ($^{\circ}$) & $250.26 \pm 0.01_{\rm stat}$ &  $250.25 \pm 0.01_{\rm stat} \pm 0.01_{\rm syst}$ & $250.26$\\
 & Dec ($^{\circ}$) &  $-46.32 \pm 0.01_{\rm stat}$ &  $-46.29 \pm 0.01_{\rm stat} \pm 0.03_{\rm syst}$ & $-46.30$	 \\
\multicolumn{2}{c}{Degree of Freedom (DoF)} & 9 & 10 & 5 \\
\multicolumn{2}{c}{TS$_{\rm ext}$ \tiny(point vs model)} & $-$ & 35 & 34 \\
\end{tabular}
\end{table*}

%Here they are added to provide a proper description of the region but their large distance with respect to our sources of interest ensure that they should not affect significantly the results derived in the following.

 \begin{table}[ht!]

\caption{Names, spectral indices and TS of the additional sources found by \textit{Fermipy} in the region of  \ang[astroang]{15}$\times$ \ang[astroang]{15} centered on HESS J1640$-$465 above 200 \si{\mega\electronvolt} ($\dagger$) and above 1 \si{\giga\electronvolt}.}
\label{table:add}
\begin{tabular}{ c c c  }

\multicolumn{3}{c}{} \\
\hline
 Name & Index  & TS \\
 \hline
PS J1617.5$-$5104$^{\dagger}$ & $\Gamma = 2.02 \pm 0.02 $ & 37 \\
PS J1708.8$-$4007$^{\dagger}$ & $\Gamma = 3.02\pm 0.02$ & 55\\
\hline
PS J1633.7$-$4755 & $\Gamma = 2.71 \pm 0.17$ & 98 \\
PS J1638.4$-$4715 & $\Gamma = 2.62 \pm 0.02$ & 33 \\
PS J1640.2$-$4803 & $\Gamma = 3.26 \pm 0.34$ & 34 \\
PS J1647.4$-$4541 & $\Gamma = 2.22 \pm 0.17$ & 32 \\

PS J1652.3$-$4433 & $\Gamma = 2.84 \pm 0.02$ & 54 \\
PS J1703.2$-$4145 & $\Gamma = 3.45 \pm 0.40$ & 59 \\

\end{tabular}

\end{table}

The spatial analysis of both sources was performed iteratively with \textit{Fermipy}. Using the updated source model, we relocalized HESS J1640$-$465 and HESS J1641$-$463 using a point-like source assumption. Then we fixed the estimated position of HESS J1641$-$463 and looked for a potential extension of HESS J1640$-$465 (and computed its position). Finally, we used the new position and extension of HESS J1640$-$465 to re-localize HESS J1641$-$463 and search for its extension.

To find the best morphological model, the different spatial shapes were tested: (A) 2 point-like sources relocalized by \textit{FermiPy}, (B) a 2D Gaussian for HESS J1640$-$465 plus a point source for HESS J1641$-$463, relocalized by \textit{Fermipy} and (C) the best positions and extension obtained by H.E.S.S. for both sources. For  all these tests, a power-law  and a log-parabola spectral shape were  assumed for HESS J1640$-$465 and  for HESS J1641$-$463 respectively.

 Table \ref{table:best_morpho} gives the positions and extensions together with the associated number of degrees of freedom and the TS$_ {\rm ext} $\footnote{TS$_ {\rm ext} $ is defined as twice the log likelihood difference between an extended model of the source and the point-like source model.} value. This is the first time that the extension of HESS J1640$-$465 is significantly measured above 1 \si{\giga\electronvolt} using the \textit{Fermi}-LAT data with a TS$_ {\rm ext} $ = 35 which corresponds to a significance of $\sim$ 6$\sigma$. The best fit position found using \textit{Fermipy} for HESS J1640$-$465 as a 2D Gaussian is $\alpha$(J2000) =  \ang[astroang]{250.17} $\pm$ \ang[astroang]{0.01}$_{\rm stat}$ $\pm$ \ang[astroang]{0.01}$_{\rm syst}$, $\delta$(J2000) =   \ang[astroang]{-46.55} $\pm$ \ang[astroang]{0.01}$_{\rm stat}$ $\pm$ \ang[astroang]{0.01}$_{\rm syst}$ with an extent of $\sigma$ = \ang[astroang]{0.053} $\pm$ \ang[astroang]{0.011}$_{\rm stat}$ $\pm$ \ang[astroang]{0.008}$_{\rm syst}$. The extension measured by H.E.S.S. of \ang[astroang]{0.072} $\pm$ \ang[astroang]{0.003} \citep{abramowski_hess_2014} is consistent with our best estimate, considering the uncertainties on both measurements.
%The best model obtained by \textit{Fermipy} is not significantly better than the H.E.S.S. extension and position values extracted from \cite{abramowski_discovery_2014} and \cite{abramowski_hess_2014} (1$\sigma$ better than H.E.S.S.).

No significant extension was found for HESS J1641$-$463 using \textit{Fermipy} (with a TS$_{\rm ext}=6$ for a 2D Gaussian, equivalent to 2.45$\sigma$).  The best-fit position for HESS J1641$-$463 as a point-like source is $\alpha$(J2000) = \ang[astroang]{250.25} $\pm$ \ang[astroang]{0.01}$_{\rm stat}$, $\delta$(J2000) = \ang[astroang]{-46.29} $\pm$ \ang[astroang]{0.01}$_{\rm stat}$. The good agreement between the estimated  positions of HESS J1640$-$465 and HESS J1641$-$463 in the GeV band  by \textit{Fermipy} (B) and in the TeV range covered by H.E.S.S. (C) tends to show that H.E.S.S. and the \textit{Fermi}-LAT detect the same sources.

We used the best positions and extension obtained by H.E.S.S. for HESS J1640$-$465 and HESS J1641$-$463 (model C) to derive their spectra, since the best model obtained by \textit{Fermipy} is not significantly better than the H.E.S.S. extension and position values extracted from \cite{abramowski_discovery_2014, abramowski_hess_2014}.

\subsection{Spectral analysis}
\label{sec:spec}
The following spectral analysis was performed using \textit{gtlike} in a summed likelihood binned analysis with a combination of the four PSF event-types between 200 \si{\mega\electronvolt} up to 1 \si{\tera\electronvolt}. The source model of the region contains the 8 additional sources presented in Table \ref{table:add}, as well as the best morphological parameters from H.E.S.S. for HESS J1641$-$463 and HESS J1640$-$465. In order to obtain the best model of the observed emission, we made a comparison between multiple spectral shapes for both sources (see Table \ref{table:spec_model}).

\begin{table*}[ht]
\centering
\caption{Comparison of spectral models (200 MeV -- 1 TeV). The left value is for HESS J1640$-$465 and the right value is for HESS J1641$-$463. The \textit{Index} column is for both the index (power-law), and for $\alpha$ (log-parabola).}
\label{table:spec_model}

\begin{tabular}{cccccc}
\multicolumn{6}{c}{} \\
\hline\hline
HESS J1640$-$465 & HESS J1641$-$463 & Index$_{\tnote{1}}$ & $\beta$ & DoF & $\Delta TS_{{modelvs\hspace{0.1cm} PLPL}} $ \\
\hline
Power-law & Power-law & 1.7$\pm$0.1/2.7$\pm$0.06 & $-$/$-$ & 4 & $-$ \\
Power-law & Log-parabola & 1.8$\pm$0.1/2.7$\pm$0.1  & $-$/0.11$\pm$0.03 & 5 & 16 \\
Log-parabola & Power-law & 1.7$\pm$0.1/2.7$\pm$0.08  & 0.0003$\pm$0.005/$-$ & 5 & 2 \\
Log-parabola & Log-parabola & 1.7$\pm$0.2/2.6$\pm$0.1  & 0.0004$\pm$0.005/0.17$\pm$0.05 & 6 & 18 \\

\hline

\end{tabular}
%\begin{tablenotes}
%\item [1] The left value is for HESS J1640$-$465 and the right value is for HESS J1641$-$463. The \textit{Index} column is for both the index (power-law), and for $\alpha$ (log-parabola).
%\end{tablenotes}

\end{table*}

Assuming the best spectral shape, the $\gamma$-ray sources observed by the LAT above 200 \si{\mega\electronvolt} are detected with a TS of 232 for HESS J1640$-$465 and 813 for HESS J1641$-$463. The spectrum of HESS J1641$-$463 is well described by a log-parabola ($ \rm{d} N/ \rm dE \propto E^{-(\alpha+\beta log (E/Eb)}) $ with $\alpha = 2.7 \pm 0.1_ {\rm stat} \pm 0.2_ {\rm syst} $ and a significant curvature (4$\sigma$ improvement with respect to a power-law model) of $\beta = 0.11 \pm 0.03_ {\rm stat} \pm 0.05_ {\rm syst}$. An integrated flux of ($4.54 \pm 0.31_{\rm stat} \pm 1.53_{\rm syst}) \times 10^{-11}$ erg cm$^{-2}$ s$^{-1}$ is obtained between 200 \si{\mega\electronvolt} and 1 \si{\tera\electronvolt}.

The spectrum of  HESS J1640$-$465 is well described by a power-law ($ {\rm d}N/{\rm d}E  \propto E^{-\Gamma}$)  with a hard spectral index $\Gamma= 1.8 \pm 0.1_{\rm stat}\pm 0.2_{\rm syst}$ and integrated flux between 200 MeV and 1 TeV of ($40.0\pm 0.5_{\rm stat}\pm 0.3_{\rm syst})\times10^{-11}$ erg cm$^{-2}$ s$^{-1}$. The spectrum does not show any significant curvature (see Table~\ref{table:spec_model}).

The results presented here are given with statistical  and systematic uncertainties. The systematic uncertainties are due to our imperfect model of the Galactic diffuse emission ($Err_ {\rm adm} $), to the spatial shape of HESS J1640$-$465 ($Err_ {\rm shape} $) and to our knowledge of the \textit{Fermi}-LAT IRFs ($Err_ {\rm irf} $). The systematic uncertainties are added in a quadrature as $Err_ {\rm syst} = \sqrt{(Err_{\rm adm})^{2}+(Err_{\rm irf})^{2}+(Err_{\rm shape})^{2}}$. 

The first uncertainty was estimated using the results of fits taking into account eight alternative interstellar diffuse emission models and comparing the results to the nominal fit, as in \cite{acero_first_2016}. The uncertainties on the morphology of HESS J1640$-$465 was estimated by fitting the $\gamma$-ray emission with the best spatial models given by \textit{Fermipy}. The third uncertainty  was estimated by using modified IRFs whose effective areas bracket the nominal ones\footnote{\textit{\scriptsize https://fermi.gsfc.nasa.gov/ssc/data/analysis/LAT\_caveats.html}]}. The same uncertainties were taken into account for HESS J1641$-$463, but without the uncertainties on the morphology $Err_ {\rm shape} $.

\begin{figure}[ht]
    
    \includegraphics[scale=0.25]{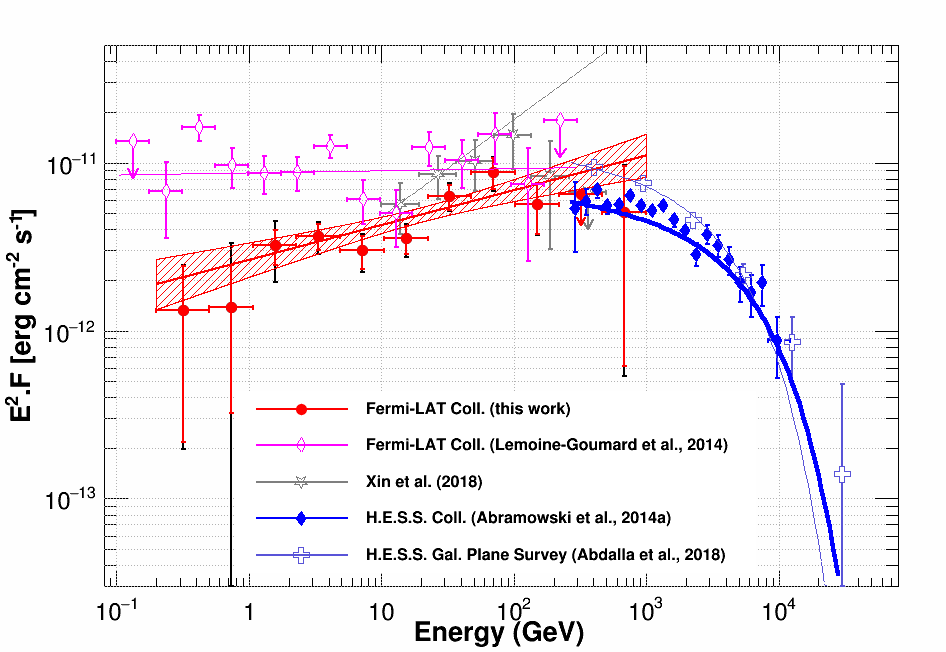}
    \caption{$\gamma$-ray spectral energy distribution (SED) of HESS J1640$-$465 between 200 \si{\mega\electronvolt} and 1 \si{\tera\electronvolt} with the \textit{Fermi}-LAT data, using the best Gaussian spatial model from H.E.S.S. The purple thin line and the open diamonds are from the latest publication by the \textit{Fermi}-LAT collaboration \citep{lemoine-goumard_hess_2014}. The grey line and open stars are from \cite{xin_hess_2018}. The red line and dashed area represent our best-fit and 68\% confidence band of the fitted \textit{Fermi}-LAT spectrum. The red data points (circle) are the flux estimated in each energy bins together with their statistic (red) and statistic  plus systematic (black) uncertainties. A 95\% confidence level upper limit (UL) is computed when the significance in an energy bin is lower than 1. The blue diamonds and line represent the spectrum from the H.E.S.S. analysis extracted from \cite{abramowski_discovery_2014}. The thin blue line and the open crosses are from the HESS Galactic Plane Survey \citep{hess_survey}}.\\
    \label{fig:sed1640}
\end{figure}
\begin{figure}[ht]
    
    \includegraphics[scale=0.25]{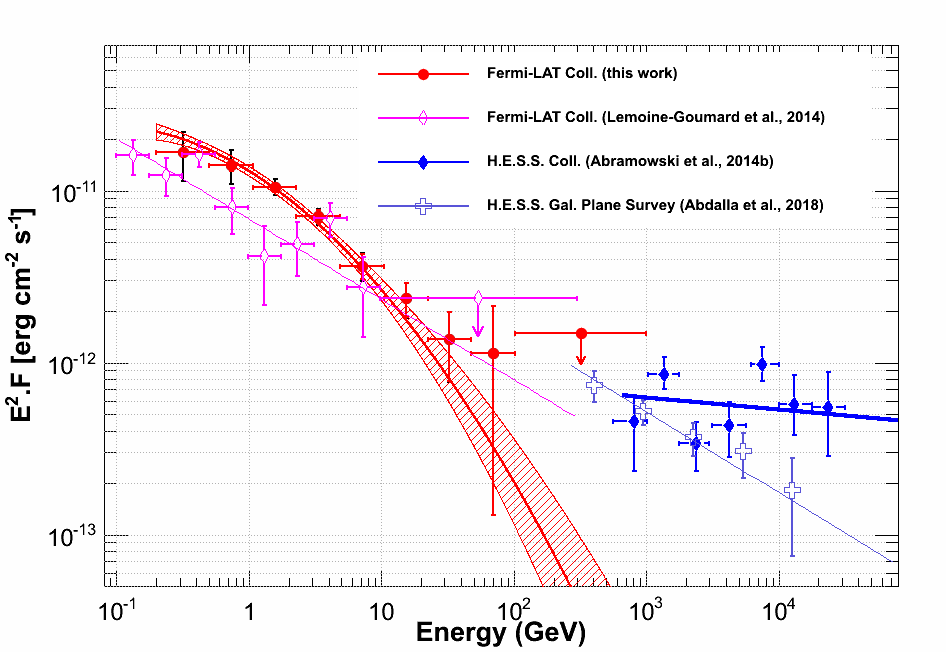}
    \caption{$\gamma$-ray spectral energy distribution of HESS J1641$-$463 between 200 \si{\mega\electronvolt} and 1 \si{\tera\electronvolt} with the \textit{Fermi}-LAT data, using the best H.E.S.S. localization, using the same conventions as in Figure 2.}
    \label{fig:sed1641}
\end{figure}

Figures \ref{fig:sed1640} and \ref{fig:sed1641} present the spectra obtained for HESS J1640$-$465 and HESS J1641$-$463 by dividing the 0.2 -- 1000 \si{\giga\electronvolt} interval into 11 logarithmically spaced energy bins. A maximum likelihood spectral analysis was performed in each interval, considering both sources as a power-law shape of index $\Gamma$ = 2. Only the prefactor of the sources within  \ang[astroang]{3.5} from HESS J1640$-$465, the normalization of the diffuse Galactic and isotropic emission were left free. When the statistical significance was below 1$\sigma$, a 95\% confidence level upper limit was computed. As we did not detect any signal with the \textit{Fermi}-LAT data in the 0.1 -- 1 \si{\tera\electronvolt} band for HESS J1641$-$463, an upper limit was obtained for the combination of the 3 last bins. Both statistical (red) and statistical plus systematic (black) uncertainties are shown on the figures \ref{fig:sed1640} and \ref{fig:sed1641}. 

 There are several differences between this analysis and the previous analysis reported in \cite{lemoine-goumard_hess_2014}: the use of 8 years of Pass 8 data, the new 4FGL source list, updated Galactic and isotropic diffuse emission models, and a curved spectral shape for HESS J1641$-$463. Furthermore, the lower energy range is different between the two analysis, which as a strong impact on the first point of the SED (from UL to flux). Two hypotheses were tested in order to recover the results of the previous analysis of 2014. The first test was to remove the curvature from HESS J1641$-$463. Doing so, the index of HESS J1640$-$465 became softer (with an index $\Gamma =1.85 $) but not enough to retrieve the index of $\sim$2.  The biggest difference between the two analysis came from the neighboring source 4FGL J1639.8$-$4642 (see Figure 1). This source is  \ang[astroang]{0.2} away from HESS J1640$-$465 and was not detected in the 3FGL source list used in the previous analysis. Removing this source from the source model of the region degraded the log-likelihood value of the fit by $\Delta$TS = 102. Not taking this source into account results in a soft spectral index for HESS J1640$-$465 of $\simeq$ 2.1. The log-parabola shape of this nearby source, with a significant curvature of $\beta=0.15 \pm 0.058$, probably contaminated the emission of HESS J1640$-$465 at low energy, yielding  the softer\footnote{compared to this analysis.} spectral index of $\sim$2  reported in the previous analysis. This was also the case in the analysis of \cite{slane_fermi_2010} with additional contamination coming from HESS J1641$-$463, separated by  \ang[astroang]{0.25} as well, giving a spectral index of $\sim$2.3 for HESS J1640$-$465.

\section{Multi-wavelength context}
\label{sec:multi}
\subsection{X-ray counterpart}
\label{sec:xray}

\begin{figure}[ht]
    \centering
    \includegraphics[scale=0.3]{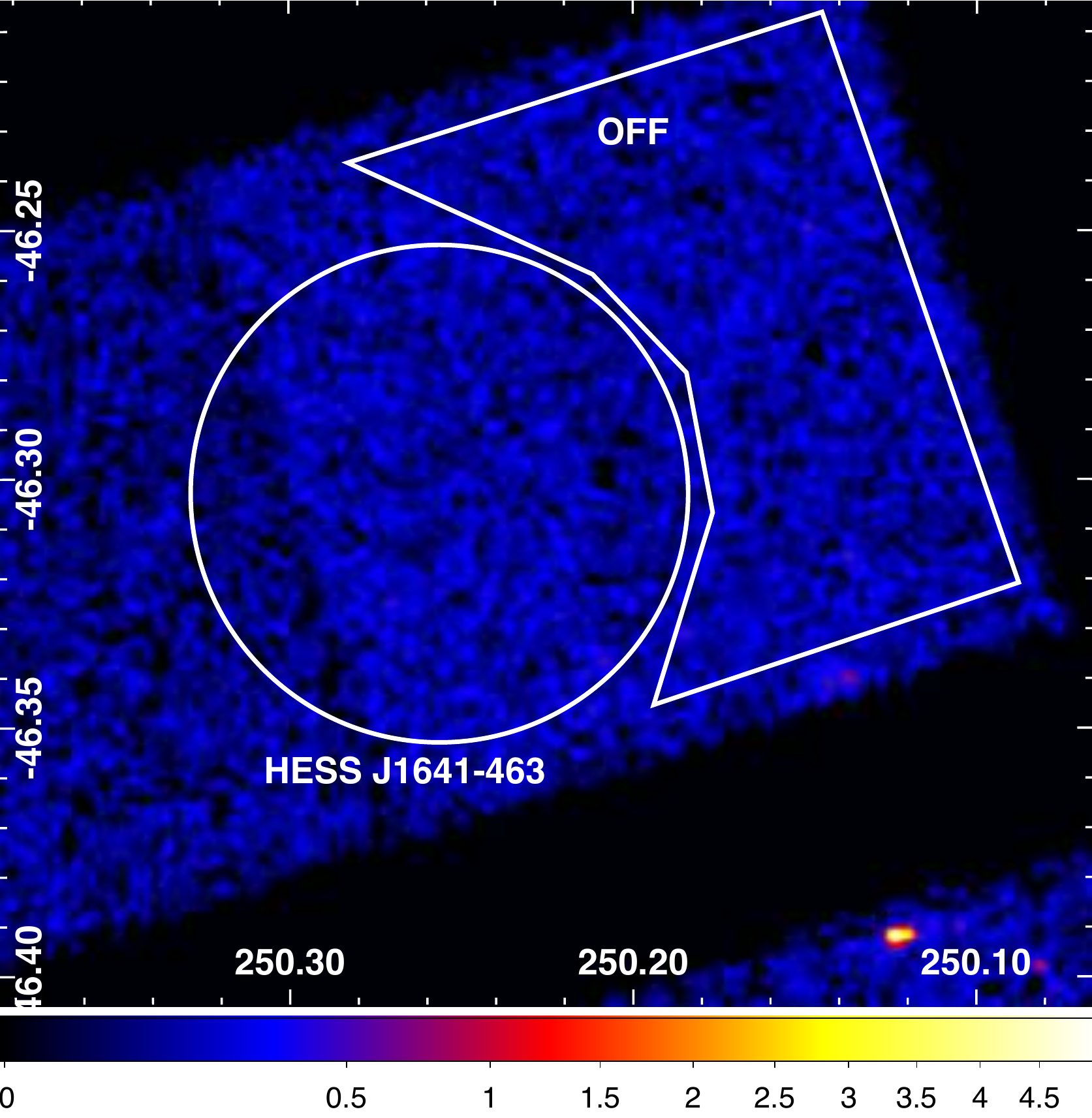}
    \caption{Counts map in the 0.5$-$6 \si{\kilo\electronvolt} energy band of the \textit{Chandra} observation smoothed with a \ang{;;6} kernel. The scale bar shows photons per pixel. The circle is centered at the best-fit position for HESS J1641$-$463, with a radius that shows the upper limit on its extent as derived in the H.E.S.S. Galactic Plane Survey. The definition of the OFF region used to estimate the astrophysical and instrumental background is also shown. }
    \label{fig:xray}
\end{figure}

The region has been observed both with \textit{XMM-Newton} and the \textit{Chandra} telescopes for follow-up observation of HESS J1640$-$465. In the \textit{XMM-Newton} observation (ObsID 0302560201; exposure of 22 \si{\kilo\second}), HESS J1641$-$463 lies in the border of the field of view and its upper limit of 3$\arcminute$ on its extent  is not fully covered by the camera. 
In the \textit{Chandra} observation (ObsID 12508; exposure of 19 \si{\kilo\second}), the source and its upper limit on its extent is fully contained in the ACIS-S3 chip and the analysis that follows is performed with the Chandra data only.

As shown in Figure \ref{fig:xray}, no significant sources are detected within the source extent, and we computed a conservative upper limit assuming an extended source based on the HESS position, $\alpha$(J2000)= 16h41m2.1s, $\delta$(J2000) =\ang{-46;18;13}, and using the H.E.S.S. upper limit of 3$\arcminute$ as the extent.
To compare with other multi-wavelength measurements, the X-ray upper limit requires an assumed value of the X-ray absorption (N$_{\rm H}$) along the line of sight. As this N$_{\rm H}$ value and the distance to the source are unknown, we estimated the upper limit  above 2 \si{\kilo\electronvolt} to limit this systematic effect as above this energy the X-ray absorption effect becomes negligible. For the upper bound, the energy range was  limited to 6 \si{\kilo\electronvolt} as  above this energy the instrumental background becomes dominant.
The astrophysical and instrumental background were derived using an OFF region (shown in Figure \ref{fig:xray}) taken from the same chip\footnote{As the S3 chip is back-illuminated, the instrumental background is slightly different and can not be estimated from the other front-illuminated chips in this observation.}.
In this restricted energy range, the OFF region is well-modeled with two power-laws. The ON region is modeled with an absorbed power-law with a spectral index fixed to 2 and the N$_{\rm H}$ value is taken to the Galactic HI value at this position ($2 \times 10^{22}$  \si{\per\square\centi\meter}). As the unabsorbed flux is estimated above 2 \si{\kilo\electronvolt}, the choice in the N$_{\rm H}$ value has a negligible impact on the flux upper limit.
With the models described above, the upper limit is derived by computing a likelihood profile on the flux jointly fitting the flux and the background from the OFF region.
The resulting 95\% confidence unabsorbed flux upper limit is $1.58\times10^{-13}$ erg \si{\per\square\centi\meter}\si{\per\second}.

\subsection{Radio counterpart}
\label{sec:radio}

The Southern Galactic Plane Survey \citep[SGPS,][]{haverkorn_southern_2006} covers G338.5$+$0.1 at 1.4~\si{\giga\hertz}, with a resolution of 100\arcsecond.  The SGPS image does not show any obvious local maximum at the position of HESS J1641$-$463. In order to constrain the level of any radio emission associated with HESS J1641$-$463, Gaussian models with varying amplitudes, each with FWHM of 3\arcminute, were added to the observed SGPS image. From these an upper limit of $\sim 2$~Jy for any radio emission was estimated.

\section{Discussion}

In this paper, we have investigated the \si{\giga\electronvolt} emission from HESS J1640$-$465 and HESS J1641$-$463.
Multiple scenarios can explain the $\gamma$-ray emission from these sources. These scenarios include the emission from accelerated particles within the SNRs coincident with each source,  molecular clouds illuminated by cosmic rays (CRs) from close-by SNR (either from the SNR G338.3$-$0.0 for HESS J1640$-$465 or the SNR G338.5$-$0.1 for HESS J1641$-$463), a PWN (detected in X-ray for HESS J1640$-$465 and hypothetical for HESS J1641$-$463) and an active galactic nucleus (AGN) for HESS J1641$-$463. These scenarios are discussed below.

\subsection{HESS J1640\texorpdfstring{$-$}{-}465}

The detection of the extension of HESS J1640$-$465 in the \si{\giga\electronvolt} energy band using the \textit{Fermi}-LAT data, together with the new spectrum, help us constrain the origin of its emission. The hard spectrum reported in this work smoothly connects with the \si{\tera\electronvolt} spectral points extracted from \cite{abramowski_hess_2014} as shown in Figure \ref{fig:sed1640}.  Previous studies of this source have interpreted its high-energy emission as being produced by high-energy leptons accelerated in a PWN (e.g., \citealt{slane_fermi_2010, abramowski_discovery_2014}) or high-energy hadrons accelerated in the associated SNRs.  In Section \ref{sec:pwn}, we analyze the properties of this source assuming a PWN origin for the observed $\gamma$-rays, while in Section \ref{sec:snr} we assume that the observed $\gamma$-rays are primarily produced as the result of interactions between high-energy protons or electrons accelerated in the SNR and the surrounding medium.

\subsubsection{Pulsar wind nebula scenario}
\label{sec:pwn}

Since the first detection of $\gamma$-ray emission from the Crab Nebula \citep{fazio72}, numerous $\gamma$-ray sources have been associated with PWNe \citep[e.g.][]{klepser_population_2013}.  In these sources, the rotational energy of a strongly magnetized, rapidly rotating neutron star (i.e., a pulsar) powers a highly-relativisic $\rm e^{\pm}$ wind (e.g., \citealt{goldreich69}).  The confinement of this ``pulsar wind'' by the surrounding medium creates a PWN (e.g., \citealt{pacini73}), inside which leptons are accelerated to high-energies (e.g., \citealt{nishikawa05, spitkovsky08}).  In such systems, $\gamma$-rays are believed to be produced through inverse Compton scattering from these energetic particles. For the clarity of this section, the details on the equations and on the modeling  were placed in the Appendix \ref{sec:pwndetails}.

When a pulsar is young it resides in the SNR created by the interaction between the material ejected in the progenitor explosion and the surrounding medium (e.g., \citealt{kennel84}).  During this time, the evolution of its PWN is sensitive to the spin-down evolution of the pulsar, the content of the pulsar wind, and the evolution of the surrounding SNR (e.g. \citealt{reynolds84}).  As a result, the properties of such a PWN can be used to determine the initial period  of the associated neutron star, the magnetization and particle spectrum of its pulsar wind, and the mass and initial kinetic energy of the supernova explosion.  Currently, this is best done by reproducing the dynamical and radiative properties of a PWN inside a SNR with a model for its evolution, as described in a recent summary by \cite{gelfand17}.  This has been successfully done for numerous systems (e.g., \citealt{chevalier05, bucciantini11, tanaka11, torres14, zhu18}) -- including HESS~J1640$-$465 using previous measurements of its $\gamma$-ray spectrum (e.g., \citealt{gotthelf_nustar_2014}).   As described in detail in Appendix \ref{sec:pwndetails}, we determined if such models can reproduce the updated $\gamma$-ray spectrum derived in Section \ref{sec:spec}.  The combination of model parameters which best fit the observed properties of this source  ($\chi^2 \approx 38$, or reduced $\chi^2 \approx 1.9$ given the 20 degrees of freedom) are given in Table \ref{tab:pwn_model_pars}, with the predicted values for the observed quantities for this combination also given in Table \ref{tab:pwn_obsprop}, in the Appendix.  As shown in Figure \ref{fig:pwn_sed}, the spectral energy distribution (SED) predicted by our model provides an excellent fit to the observed quantities, especially its $\gamma$-ray emission.

\begin{table}[tbh]
    
    \caption{``Best-fit'' Model Parameters for HESS J1640$-$465.}
    \label{tab:pwn_model_pars}
    
    {\setlength{\tabcolsep}{2.1pt}\begin{tabular}{cc}
    \hline
    \hline
    {\sc Parameter} & {\sc Value} \\
    Supernova Explosion Energy $E_{\rm sn}$ & $1.4\times10^{51}~{\rm erg}$  \\
    Supernova Ejecta Mass $M_{\rm ej}$ & $11~{\rm M}_\odot$ \\
    ISM density $n_{\rm ism}$ & $0.009~{\rm cm}^{-3}$ \\
     Spin-down timescale $\tau_{\rm sd}$ & $4.15~{\rm years}$ \\
    Magnetization $\eta_{\rm B}$ & 0.10 \\
    Minimum energy of injected $\rm e^{\pm}$, $E_{\rm min}$ & 270~GeV \\
    Break energy of injected $\rm e^{\pm}$, $E_{\rm break}$ & 16~TeV \\
    Maximum energy of injected $\rm e^{\pm}$, $E_{\rm max}$ & 0.3~PeV \\
    Low-energy index of injected $\rm e^{\pm}$ $p_1$ &  1.94 \\
    High-energy index of injected $\rm e^{\pm}$ $p_2$ &  2.94 \\
    Number of added IC photon field & $\equiv 2$ \\
    Temperature of added photon fields & 5.4~K, 27000~K \\
    Norm of added photon fields & 0.16, $5.2\times10^{-13}$ \\
    Distance $d$ & 12.1~\si{\kilo\parsec} \\
    \hline
    \hline
    \end{tabular}}
    \tablecomments{The observed properties of HESS J1640$-$465 used in this fit are listed in Table \ref{tab:pwn_obsprop}.}
\end{table}

\begin{figure}[tbh]
    
    \includegraphics[scale=0.48]{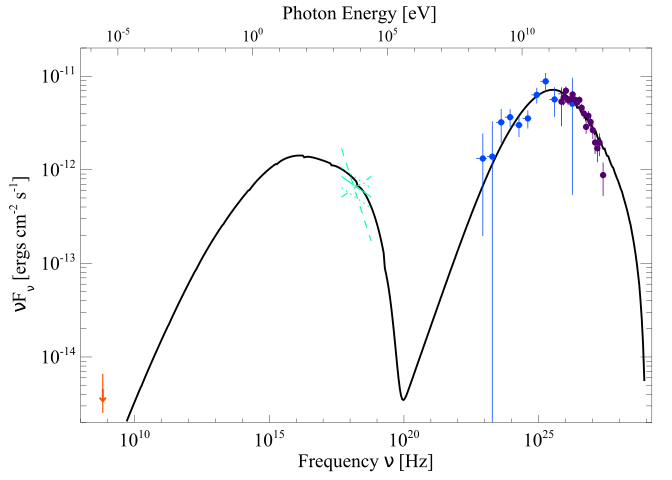}
    \caption{Observed (color points) and predicted (black line) SED of the PWN in HESS J1640$-$465.  The model parameters used in the prediction are listed in Table \ref{tab:pwn_model_pars}.}
    \label{fig:pwn_sed}
\end{figure}

While our model produces a statistically good fit to the observed properties of this source, determining if it is plausible requires considering the resultant physical properties, as listed in Tables \ref{tab:pwn_model_pars} and \ref{tab:pwn_model_pred}.  While our fit suggests the mass ($M_{\rm ej} \approx 10~{\rm M}_\odot$) and initial kinetic energy ($E_{\rm sn} \approx 10^{51}~{\rm ergs}$) of the supernova ejecta are close to ``typical'' values \citep[e.g.][]{baade34}, it requires that the resultant pulsar (PSR~J1640$-$4631) has an extremely short spin-down time-scale \footnote{$\tau_{\rm sd}$ is the spin-down timescale, equivalent to the ``characteristic age'' of the pulsar at birth.}: $\tau_{\rm sd} \approx 4~{\rm years}$ instead of $\tau_{\rm sd} \sim 1000~{\rm years}$ inferred for other pulsars associated with PWNe
%\footnote{$\tau_{\rm sd}$ is the characteristic timescale in which the pulsar loses a fraction of its rotational energy;} 
(cf Appendix \ref{sec:pwndetails}) (e.g., \citealt{bucciantini11, tanaka11, torres14, gelfand14, gelfand15, zhu18}).  If correct, then the pulsar was born spinning very rapidly, with an initial spin-period (e.g., \citealt{goldreich69,pacini73,gaensler_evolution_2006}):
\begin{eqnarray}
\label{eqn:p0}
P_0 & = & P \left(1+\frac{t_{\rm age}}{\tau_{\rm sd}} \right)^{-\frac{1}{p-1}} \approx 9.5~{\rm ms},
\end{eqnarray}
for a current period $P \approx 206~{\rm ms}$ \citep{gotthelf_nustar_2014}.The initial period-derivative of this pulsar $\dot{P}_0$ predicted by this solution is:
\begin{eqnarray}
\dot{P}(t) & = & \frac{P_0}{\tau_{\rm sd}(p-1)} \left(1+\frac{t}{\tau_{\rm sd}}\right)^{\frac{2-p}{p-1}} \\
\dot{P}_0 & = & \frac{P_0}{\tau_{\rm sd}(p-1)} \approx 3.38 \times10^{-11}~\frac{\rm s}{\rm s},
\end{eqnarray}
which suggests as initial surface dipole magnetic field strength $B_{s0} \equiv 3.2\times10^{19}\sqrt{P_{0} \dot{P}_{0}} \approx 1.8\times10^{13}~{\rm G}$ -- only $\sim30\%$ larger than the $B_{s} \approx 1.4\times10^{13}~{\rm G}$ inferred from its current spin-down properties \citep{gotthelf_nustar_2014}.  However, this model requires this pulsar had an extremely high initial spin-down luminosity $\dot{E}_{0} \approx 1.55\times10^{42}$~erg~\si{\per\second}, $\sim100\times$ higher than that estimated in analyses of other PWNe (e.g., \citealt{torres14}).  However, the total rotational energy injected into the SNR at early times, ${\dot{E}}_{0_{\tau{\rm sd}}} \approx 2.0\times10^{50}~{\rm erg}$, is considerably lower than the $\gtrsim10^{51}~{\rm erg}$ explosion energy.  Lastly, the magnetization of the wind produced by this pulsar $\eta_{\rm B} \sim 10^{-1}$ is significantly larger than the magnetization $\eta_{\rm B} \sim 10^{-3}$ inferred from many other pulsars (e.g., \citealt{bucciantini11,tanaka11,torres14,gelfand15,zhu18}), but is comparable to the $\eta_{\rm B} \sim 0.04$ inferred by a similar analysis of Kes~75 \citep{gelfand14}, whose associated neutron star also has $B_{s} \gtrsim10^{13}~{\rm G}$.  It is also worth noting that PSR J1640-4631 is one of the few pulsars with a measured braking index, 3.15 $\pm$ 0.03, and the first one with p$>$3, the canonical value for pure magnetic dipole radiation \citep{archibald16},  measured from a 2.3 year long phase-connected timing solution.  As such, it is perhaps not surprising that it would have an initial spin period $P_{0}$ and initial spin-down luminosity $\dot{E}_{0}$  faster and higher, respectively, than most neutron stars.

\begin{table}[tbh]
\begin{center}
\caption{Physical Properties of HESS J1640$-$465 predicted by our model with the parameters listed in Table \ref{tab:pwn_model_pars}.}
\label{tab:pwn_model_pred} 
   {\setlength{\tabcolsep}{2.1pt} \begin{tabular}{cc}
    \hline  
    \hline
    True age $t_{\rm age}$ & 3094~years \\
    Initial Spin Period $P_0$ & 9.5~{\rm ms} \\
    Initial Period Derivative $\dot{P}_0$ & $3.38\times10^{-11}~\frac{\rm s}{\rm s}$ \\
    Initial Surface Magnetic Field $B_{s0}$ & $1.8\times10^{13}$~G \\
    Initial Spin-down Luminosity $\dot{E}_0$ & $1.5\times10^{42}~\frac{\rm erg}{\rm s}$ \\
    Energy Density of added photon fields & 0.64~$\frac{\rm eV}{\rm cm^3}$, 1.3~$\frac{\rm keV}{\rm cm^3}$\\
    \hline
    \hline
    \end{tabular}}
\end{center}

\end{table}

In addition, our model requires that this SNR be expanding into an environment with an extremely intense UV photon field, suggesting it is near a collection of massive stars capable of producing the $T\sim25000~{\rm K}$ photon field required by our model to reproduce the GeV emission from this source (Table \ref{tab:pwn_model_pars}).  In fact, a cluster of young massive stars Mc81 is $\sim8^\prime$ away from HESS J1640$-$465 \citep{davies12}, with a distance ($d=11\pm2~{\rm kpc}$) and high-extinction ($A_V \sim 45$) which are consistent with the distance $d \sim 12$~kpc favored by our modeling of the PWN (Table \ref{tab:pwn_model_pars}) and the considerable $N_H \sim (1-2)\times10^{23}~{\rm cm}^{-2}$ along the line of sight to this source inferred from its X-ray spectrum \citep{gotthelf_nustar_2014}. Since the projected distance between this stellar cluster and HESS J1640$-$465 is $\sim30~{\rm pc}$, this stellar cluster is close enough to provide the intense photon field our modeling required to reproduce the observed $\gamma$-ray spectrum.

If the progenitor of HESS J1640$-$465 was originally a member of this cluster but dynamically ejected before it exploded, as suggested by \cite{davies12}, it likely was its most massive member since it was the first to explode.  In this case, its initial mass should have been $\gtrsim65~{\rm M}_\odot$ inferred for a Wolf-Rayet star in this cluster (Mc81-3; \citealt{davies12}).  If correct, the $M_{\rm ej}\sim11~{\rm M}_\odot$ of material ejected in this explosion (Table \ref{tab:pwn_model_pars}) suggests the progenitor lost most of its mass before exploding.  In this case, its powerful stellar winds would create a low-density bubble into which the ejecta initially expands.  This is supported by the low ISM density ($n_{\rm ism} \sim 0.01~{\rm cm}^{-3}$) preferred by our modeling (Table \ref{tab:pwn_model_pars}).  While seemingly at odds with the dense molecular material detected nearby \citep[e.g.][]{lau2017}, the short lifetime of such massive stars suggests they explode in the vicinity of their birth molecular cloud -- consistent with the suggested interaction between the SNR shell and surrounding molecular material.

\subsubsection{Pulsar scenario}

The significant extension detected by our analysis above 1 GeV, in agreement with that previously reported by the H.E.S.S. collaboration, together with our hard spectral index of 1.8 strongly disfavour the pulsar hypothesis for HESS J1640$-$465. The fact that no $\gamma$-ray pulsations have been detected coming from an X-ray pulsar with such a high $\dot{E}$ ($4.4\times10^{36}~$erg s$^{-1}$) could be explained as in \cite{johnston_galactic_2020}. These X-ray pulsars have a selection bias making them at great distance, deep in the Galactic plane, and thus below LAT sensitivity. Furthermore, this pulsar is co-located with HESS J1640$-$465 and HESS J1641$-$463, making the detection of any signal from this pulsar even more difficult.
On top of that, their geometry may be particularly unfavorable. Finally, an X-ray timing solution of 10 or 12 years is necessary for the faintest pulsars \citep{smith_searching_2019}, which is far from the 2.3 years timing solution used for PSR J1640$-$4631.

\subsubsection{SNR scenario}
\label{sec:snr}

Now, we consider the hypothesis that the $\gamma$-ray emission detected from HESS J1640$-$465 is produced by particles accelerated at the shocks of the SNR G338.3$-$0.0. Since a SNR origin for such emission was proposed by a number of authors \citep{abramowski_hess_2014,abramowski_discovery_2014,tang_self-consistent_2015,lau2017,supan2016}, we limit the discussion to few essential issues that emerged from the new observational results presented in this paper.

In the \si{\giga\electronvolt} domain, the $\gamma$-ray spectrum of HESS J1640$-$465 can be described by a power-law with slope 1.8, which is significantly harder than what was previously reported \citep{lemoine-goumard_hess_2014}.
High-energy $\gamma$-ray spectra harder than $E^{-2}$ are often interpreted as the result of inverse Compton scattering emission from electrons accelerated at the SNR shock.
This is the case, for example, for the SNR RX J1713.7$-$3946, \citep{abdo2011}, whose TeV spectrum very closely resembles that of HESS J1640$-$465 \citep{abramowski_hess_2014}.
An estimate of the energy required in accelerated electrons can be obtained by analogy with RX J1713.7$-$3946.
\citet{porter2006} showed that a leptonic (inverse Compton) fit to $\gamma$-ray data from RX J1713.7$-$3946 would require an energy of $\approx 5 \times10^{47}$ erg in form of accelerated electrons. This is a small fraction ($\approx 2 \times 10^{-3}$) of the typical supernova explosion energy ($\sim 10^{51}$ erg).
The \si{\tera\electronvolt} flux of HESS J1640$-$465 is a factor of $\lesssim 10$ smaller than that of RX J1713.7$-$3946, but its distance is 10 times larger. This implies that a leptonic interpretation of the $\gamma$-ray spectrum of HESS J1640$-$465 would require a total energy in relativistic electrons equal to $\gtrsim 5 \times 10^{48}$ erg. In other words, if the SNR shock converts the canonical $\sim 10$ \% fraction of the explosion energy into CR {\it protons}, the leptonic interpretation {\bf would require an electron-to-proton ratio} of the order of $K_{\rm ep} \gtrsim 5 \times 10^{-2}$.
%The similarity between the two spectra suggests that a leptonic scenario for HESS J1640-465 might suffer from the same problems that have been pointed out for RX J1713.7-3946.
%However, though not excluded, a leptonic interpretation would require a quite tight energy budget for CR electrons. 
Such a large value of the electron-to-proton ratio, though not excluded, is significantly larger than what is usually assumed in modelling the broadband emission from SNRs, i.e. $K_{ep} \approx 10^{-5} ... 10^{-2}$ \citep[see e.g.][]{cristofari2013}. A similarly large value of $K_{\rm ep} \approx 0.1$ was also found by \citet{abramowski_hess_2014} (see their Fig.~5, where they fit the multi-wavelength spectrum adopting a time dependent single zone model).

On the other hand, as pointed out by \citet{zirakashvili2010} and \cite{inoue2012}, an alternative, hadronic scenario may also be invoked to explain the hard $\gamma$-ray spectra which are observed from a number of SNRs.
This is because, if the SNR shock expands in a clumpy medium, propagation effects can shape the $\gamma$-ray spectrum. The reason is that, due to the finite age of the SNR, only CRs of sufficiently large energy  have enough time to diffusively penetrate into dense clumps, interact with the gas and produce gamma rays from neutral pion decay. The exclusion of lower energy CRs from the clumps will result in a very hard $\gamma$-ray spectrum \citep{gabici2014,celli2019}, that might easily match the observed spectrum of HESS J1640$-$465.
Note that a more conventional value for the electron-to-proton ratio $K_{ep} \approx 10^{-2}$ was adopted in the hadronic scenario described in \citet{abramowski_hess_2014}.

\subsection{HESS J1641\texorpdfstring{$-$}{-}463}

A possibility is that HESS J1641$-$463 is an active galactic nucleus (AGN), given its point-like morphology and curved spectrum seen in the GeV energy band.
If it is a blazar, the spectral index of 2.7 (for a log-parabola spectrum) indicates that it should be a Flat Spectrum Radio Quasar (FSRQ). With a TS of 316 in the 4FGL catalog, a FSRQ will have a probability  $>0.95$ to be seen as variable in the 4FGL and a probability of almost 90\% to present a power-law spectrum instead of a log-parabola spectrum \citep[see Figures 8 and 12 in][]{ajello_fourth_2020}. In the second release of the 4FGL catalog, the variability index\footnote{Parameter constructed from the value of the likelihood in the null hypothesis, i.e the flux of the source is constant over the time, and the value under the hypothesis where the flux in each bin is optimized.} is $\sim 7.4$. Therefore, no variability is detected in the 4FGL-DR2 using 10 years of data. Whereas the FSRQ hypothesis cannot be ruled out, it is unlikely. Instead we consider SNR, pulsar and PWN hypotheses for HESS J1641$-$463.

\subsubsection{Hadronic scenario (SNR or illuminated molecular cloud scenario)}

Interpreting the $\gamma$-ray spectrum of HESS J1641$-$463 is a more difficult task than for HESS J1640$-$465, because it is unusual in having a curved spectrum at \si{\giga\electronvolt} energies, and a hard spectrum at \si{\tera\electronvolt} energies. The curved spectrum observed by \textit{Fermi} could result either from the direct acceleration of CR protons at the shock of an aged SNR \citep{lau2017}, or from the reacceleration of ambient CRs, if the shock interacts with a cloud \citep[i.e cloud crushed model, see][]{uchiyama2010}. 
In the latter case, the shock becomes radiative and compresses (crushes)  simultaneously the gas and the accelerated cosmic rays. 
\begin{figure}[ht]
    
    \includegraphics[scale=0.25]{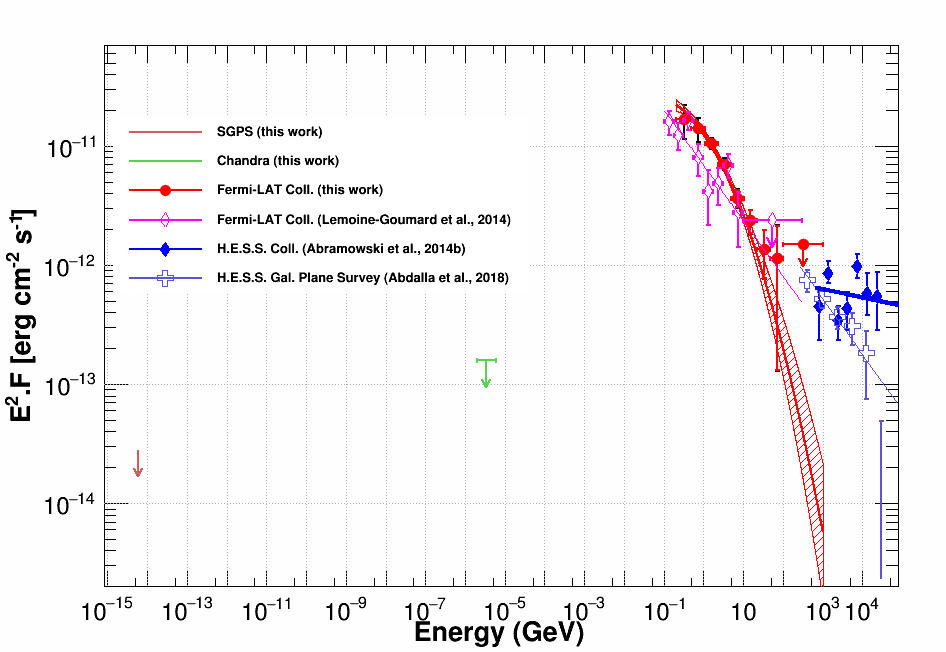}
    \caption{Multi-wavelength SED and $\gamma$-ray spectrum of HESS J1641$-$463 between 200 \si{\mega\electronvolt} and 1 \si{\tera\electronvolt} with the \textit{Fermi}-LAT data, using the best H.E.S.S. localization, using the same conventions as in Figure 3. The estimated 95\% confidence X-ray flux upper limit between 2 keV and 6 keV and the radio flux upper limit at 1.4 \si{\giga\hertz} (see Sections \ref{sec:xray} and \ref{sec:radio} for more details) are also shown.}
    \label{fig:multi}
\end{figure}
As a result, both the $\gamma$-ray and radio emission are enhanced significantly. 
However, the very large ratio between $\gamma$-ray and radio emission (see radio upper limit in Figure \ref{fig:multi}) casts serious doubts on the viability of such a scenario. Concerning an hadronic origin of the emission, the ratio between the radio flux upper limit and the GeV flux is in agreement with what is observed for other SNRs \citep[Figure 12 in][]{acero_first_2016}. 
But the X-ray upper limit also constraints severely the scenario of direct acceleration at the SNR shock. %, requiring a very low electron-to-proton ratio for accelerated particles.
Indeed, the large ratio between the X-ray upper limit and the \si{\giga\electronvolt} flux (\si{\giga\electronvolt}/X-ray$\sim 100$) is problematic for a hadronic emission. This is because, due to the large gas density in the region, a bright thermal X-ray emission would be expected \citep{katz_which_2008}.
Such an emission is not observed here.
For these reasons, a non-SNR origin of the GeV emission remains a very plausible hypothesis.

Finally, the hard \si{\tera\electronvolt} spectrum shown by HESS J1641$-$463 has been often interpreted as the result of the illumination of ambient dense gas by CRs that escaped from the SNR G338.3$-$0.0 \citep{abramowski_discovery_2014,tang_self-consistent_2015,lau2017,dewilt2017}. 
In order to test the validity of this picture, a detailed knowledge of the three dimensional distribution of the gas in the region is needed \citep[see e.g.][]{pedaletti2015}.
In this context, \citet{dewilt2017} performed an accurate study of various gas tracers and concluded that, though possible, the illumination scenario would imply that the peak of the \si{\tera\electronvolt} $\gamma$-ray emission should correlate with the peak observed in the gas distribution, which is not observed.
Another prediction of the illumination scenario is the presence of a faint X-ray emission, due to the synchrotron emission from the secondary electrons produced in the interactions of the runaway cosmic rays with the dense gas \citep{gabici2009}. However, this is expected to be faint and could be absorbed. Large ratios between the $\gamma$-ray and X-ray fluxes are indeed expected, independently on the choice of model parameters \cite[see discussion in][]{gabici2009}, and thus this scenario is consistent with the X-ray upper limit reported here (see Figure \ref{fig:multi}). In this case, the \si{\giga\electronvolt} spectrum of HESS J1641$-$463 would remain unexplained.

\subsubsection{Pulsar scenario}

The curved \si{\giga\electronvolt} spectrum of HESS J1641$-$463 reported in Section \ref{sec:spec} is similar to other $\gamma$-ray pulsars seen with the \textit{Fermi}-LAT \citep{abdo_second_2013}. We use the spectral parameters obtained here to calculate the $\gamma$-ray luminosity of the source in the GeV energy band assuming it is a pulsar.

We first attempt to fit the spectrum of HESS J1641$-$463 using a power law with a super exponential cutoff (${\rm d}N/{\rm d}E \propto E^{-\Gamma_{1}}\times exp(-(E/E_{\rm cutoff})^{\Gamma_{2}})$) generally used to characterize pulsar spectra. The best fit yields $\Gamma_{1}=2.1 \pm 0.1$,  $\Gamma_{2}=0.34 \pm 0.12$ and $E_{\rm cutoff}$ = $573.4 \pm 15.1$ \si{\mega\electronvolt}. Using the value of the  energy flux in the 0.1 to 100 \si{\giga\electronvolt} band (G$_{100}$), we can estimate the luminosity of the pulsar candidate, as done in the \textit{2$^{nd}$} \textit{Fermi}-LAT pulsar catalog \citep[hereafter referred to as \textit{2PC};][]{abdo_second_2013}:

\begin{equation}
    L_{\rm \gamma, 2PC} = 4\pi d^{2}f_{\Omega} G_{100}
    \label{eq:L_psrcat}
\end{equation} 
\vspace{0.3cm}
where $d$ is the distance of the source in \si{\kilo\parsec} (estimated to 11 \si{\kilo\parsec}) and $f_{\Omega}$ is the beam correction factor \citep{romani_constraining_2010}.
We assume $f_{\Omega} \approx$ 1 as in \textit{2PC}, which corresponds to an outer magnetosphere fan-like beam  sweeping  the  entire  sky. The resulting luminosity is $L_{\rm \gamma, 2PC}$ = 9 $ \times 10^{35} (\frac{d}{11 {\rm kpc}})^{2}$ erg~\si{\per\second}.

 The ``heuristic'' $\gamma$-ray pulsar  luminosity is defined in the \textit{2PC} as:
 
 \begin{equation}
     L^{h}_{\rm \gamma} = \sqrt{10^{33} \dot{E} }\hspace{0.1cm} {\rm erg} \hspace{0.1cm }{\rm s}^{-1}
 \end{equation} 

\begin{figure}[ht]
   
     \includegraphics[scale=0.3]{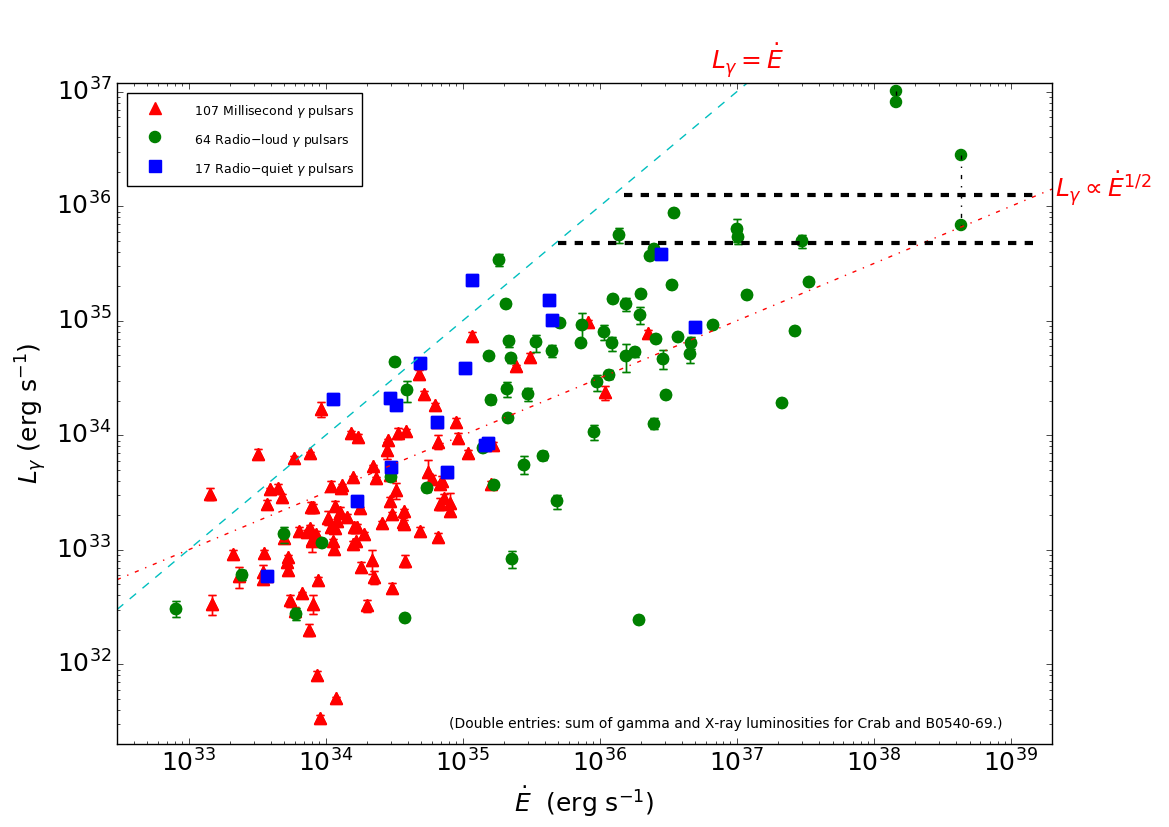}
     \caption{$\gamma$-ray luminosity $L_{\gamma} = 4\pi f_{\Omega}d^{2}G_{100}$ in the 0.1 to 100 GeV energy band
versus spin-down power $\dot{E}$ \citep[taken from ][]{abdo_second_2013}. The upper diagonal line indicates 100\% conversion of spin-down power
into $\gamma$-ray flux: for pulsars above this line, the distance $d$ may be smaller, and/or the
assumed beam correction $f_{\Omega} \equiv 1$ is wrong. The lower diagonal line indicates the heuristic
luminosity $L^{h}_{\gamma} = \sqrt{10^{33} \dot{E} }\hspace{0.1cm}$ erg s$^{-1}$. The black dash lines are the dispersion values of the spin-down power of HESS J1641$-$463  based on the associated calculated luminosity for a distance of 8 (lower) and 13 \si{\kilo\parsec} (upper).
}
     \label{fig:Edot}
 \end{figure}

As the $\gamma$-ray luminosity cannot exceed the available power of the pulsar, the spin-down power cannot be lower than the estimated luminosity of $\sim $ 9 $ \times 10^{35}$ erg~\si{\per\second}. Following the maximum value of the spin-down power derived from the heuristic luminosity lower diagonal in Figure \ref{fig:Edot}), we can say that the spin-down power of HESS J1641$-$463 is between  $\sim 10^{35}$ and $\sim 10^{39}$ erg
~\si{\per\second} for distances between 8 and 13 \si{\kilo\parsec} (see diagonals in Figure \ref{fig:Edot}).
 
As shown in Figure \ref{fig:Edot}, the $\gamma$-ray luminosity is of the same order of magnitude as that of the Crab pulsar, while the spin-down power $\dot{E}$ could be either like the Crab or a few orders of magnitude lower ($L_{\gamma,\rm  Crab}$ = 1.96 $\times 10^{35}$ erg~\si{\per\second} and $\dot{E}_{\rm Crab}$ = 4.5$ \times 10^{38}$ erg~\si{\per\second}).

%There are 15 known pulsars with $\dot{E}\geqslant10^{37}$erg s$^{-1}$ and only 3 with $\dot{E} \geqslant 1 \times 10^{38}$erg s$^{-1}$. Over the 15, ten of them have a radio flux at 1400 MHz lower than 100 $\mu$Jy (radio faint/quiet sources)(Pulsar numbers from ATNF database \footnote{https://www.atnf.csiro.au/research/pulsar/psrcat/expert.html} (Manchester et al)).
%In general, for an spin-down power above $10^{35}$erg s$^{-1}$, two third of the pulsar are seen in gamma \citep{laffon_new_2015}.\\

%Only a young pulsar could have the upper extremum value of the $\dot{E}$ interval, which could be in agreement with the lower boundary of the estimate age of the SNR (1.1 kyr - 3.5 kyr). If it is a young pulsar, it could have a strong magnetic field and should radiate in X-ray. 
%Unfortunately, as show in \S\ref{sec:xray} , only an upper limits on the X-ray flux could be derived from this source.

To test this pulsar hypothesis, we performed a search for
$\gamma$-ray pulsations using 11 years of \textit{Fermi}-LAT data. This blind  search used the two-stage approach described by
\citet{Pletsch2014+Methods}, and employed in \citet{Clark2017+FGRP4}, in which
an efficient ``semicoherent'' search is performed to scan the entire parameter
space for candidate signals, which are then followed-up using a more sensitive
fully coherent search. The lag-window used for the semicoherent stage had width
$\tau = 2^{22}\, \rm s \approx 49\,\rm d$. To increase the sensitivity of the search, we weighted the contribution of each photon to the pulsation test statistic \citep{Kerr2011}. These weights represent
the probability of each photon being emitted by the target source as opposed to
being from a nearby source or from the diffuse background. The weights were
calculated using \texttt{gtsrcprob}, using the best-fitting spectral model for
the region of interest reported in Section \ref{sec:data}. We searched for signals with frequency $\nu < 48\,{\rm Hz}$, spin-down rate $\dot{\nu} > -10^{-10}\,{\rm Hz}$ ${\rm s}^{-1}$, and sky positions within a conservative $3\sigma$ range around the best-fitting position reported in Table~\ref{table:best_morpho}. No significant candidate signals were detected. We note that the GeV spectrum of HESS~J1641$-$463 falls off significantly more slowly at high-energies than is typical for a $\gamma$-ray pulsar ($\Gamma_2 = 0.306$ vs. $\Gamma_2 = 0.667$ assumed for pulsars in 4FGL). If a pulsar is present within this source, it is therefore likely to only contribute a fraction of the total emission, with additional contributions, especially at higher energies, from a putative PWN.

The lack of detected pulsations allows us to constrain this fraction, although
this requires making several restrictive assumptions on the pulsar's spin-down
behavior that may not hold.  Using the sensitivity estimation method described
in \cite{Clark2017+FGRP4}, we estimate a 95\% confidence upper limit on the
pulsed photon flux fraction from HESS~J1641$-$463 of $42$\%. This estimate
assumes that the putative pulsar spins down at a constant rate, and does not
exhibit a significant long-term second frequency derivative, strong timing noise
or glitches, as these cause a significant loss of sensitivity for our search. Unfortunately, for a young pulsar of similar age to the SNR, these phenomena are likely to be present in the spin history over the 11 year data set searched here. The maximum frequency and spin-down rates searched here were chosen with this in mind. The limiting values correspond to a maximum spin-down power of $\dot{E} \sim 2\times10^{38}\,{\rm erg}$ ${\rm s}^{-1}$. Assuming magnetic dipole braking (which has a braking index $n=3$ where $\dot{\nu} \propto \nu^{n}$), such an energetic pulsar would have an expected spin-frequency second derivative of $\ddot{\nu} = n\,\dot{\nu}^2 / \nu = 6\times10^{-22}\,{\rm Hz}$ ${ \rm s}^{-2}$, while the semicoherent search method employed here loses sensitivity for $\ddot{\nu} \gtrsim 10^{-22}\,{\rm Hz}$ ${\rm s}^{-2}$, making it unproductive to search for higher values of $\nu$ or $\dot{\nu}$.

\subsubsection{PWN scenario}

The connection between the curved GeV and the hard TeV spectra remains unclear. This suggests that two mechanisms are producing the radiation in each energy band.

In the hypothesis where the GeV emission is produced by a pulsar, the TeV emission of HESS J1641$-$463 could be explained as its PWN emission via inverse Compton (IC) scattering. 

First, we can use the integrated energy flux between 1 and 10 TeV from the H.E.S.S. Galactic Plane Survey \citep{hess_survey} to see if the putative pulsar could generate a PWN. The TeV luminosity of HESS J1641$-$463 obtained using Equation \ref{eq:L_psrcat} is $L_{\gamma, \rm HGPS}$ = 1.05$
\times 10^{34}$ erg s$^{-1}$ for a distance of 11 kpc. This luminosity can easily be achieved using the pulsar spin-down power interval calculated in the pulsar scenario described above. This is consistent with its pre-selection as a PWN candidate in \cite{collaboration_population_2018}.

However, the upper value of the $\dot{E}$ interval obtained in Figure \ref{fig:Edot}, which could be in agreement with the lower boundary of the estimate age of the SNR (1.1 -- 3.5 kyr), would be only valid for a young pulsar. In such a case, we would expect to detect an associated PWN at radio or X-ray wavelengths. Unfortunately, as reported in Sections \ref{sec:xray} and \ref{sec:radio} , only upper limits on the X-ray and radio fluxes could be derived from this source. This is similar to the case of the relic PWN HESS J1303$-$631 \citep{beilicke_detection_2005} or HESS J1825$-$137 \citep{grondin_detection_2011}. The ratio of the synchrotron and IC fluxes mainly depends on the magnetic field of the source assuming that both emission come from the same region (one-zone model). Therefore, we use the formula in \cite{aharonian_discovery_2009} where the ratio between the $\gamma$-ray flux between 1 and 10 TeV and the X-ray upper limit between 2 and 6 keV is used to estimate an upper limit on the magnetic field of the source of $\sim$4.3 $\mu$G. This magnetic field value is similar to relic PWNe and would mean that the associated pulsar would be older than 3.5 kyr.\\

\section{Conclusion}

With more than eight years of \emph{Fermi}-LAT data, our morphological and spectral analysis allows us to bring further constrains on the origin of the $\gamma$-ray emission of the neighboring sources HESS J1640$-$465 and HESS J1641$-$463. The morphological analysis of both sources above 1 \si{\giga\electronvolt} reveal a significant extension for HESS J1640$-$465 in the \si{\giga\electronvolt} energy band, together with a hard spectrum following a power-law with a slope of $\Gamma = 1.8\pm0.1_{\rm stat}\pm0.2_{\rm syst}$, that connects well with the H.E.S.S. spectrum. When compared with the analysis of \cite{lemoine-goumard_hess_2014}, this analysis shows the strong impact of the confusion of sources on the spectral index result (see Section \ref{sec:spec}). Two origins of the multi-wavelength emission are investigated: a PWN origin associated with the X-ray pulsar PSR J1640$-$4631 correlated with the radio SNR G338.3$-$0.0, and a scenario in which protons or electrons are accelerated in the SNR shock. Both hypotheses can be used to reproduce the detected emission, though the scenario involving electrons accelerated at the SNR shock requires a high electron to proton ratio. 
The H.E.S.S. PeVatron's candidate HESS J1641$-$463 is detected as a point-like source and its \si{\giga\electronvolt} spectrum is similar to those seen from pulsars detected by Fermi-LAT. Upper limits at X-ray and radio wavelengths were also derived using \textit{Chandra X-ray} and SGPS radio observations. The new data pose severe constraints on the scenarios investigated: assuming that a pulsar produces the GeV emission detected by the LAT, constraints on its spin-down power are obtained, as well as an upper limit on the magnetic field of its putative PWN detected at TeV energies. Alternative possibilities are that protons are accelerated up to hundreds of TeV by direct acceleration of CR protons by the SNR G338.5$+$0.1 (HESS J1641$-$463) or by re-acceleration of ambient CRs cannot be discarded. However, the very large ratio between the radio upper limit and the $\gamma$-ray emission casts doubts on the viability of such scenarios. Finally, the hard TeV spectrum detected by H.E.S.S. can be reproduced by the illumination of ambient dense gas by CRs that escaped from the SNR G338.3$-$0.0 coincident with HESS J1640$-$465. Future radio (pulsation search) and $\gamma$-ray observations (morphology and spectral curvature) by CTA are essential to disentangle the nature of these two intriguing sources and very-high-energy emitters.

\section{Acknowledgements}

The \textit{Fermi}-LAT Collaboration acknowledges support for LAT development, operation and data analysis from NASA and DOE (United States), CEA/Irfu and IN2P3/CNRS (France), ASI and INFN (Italy), MEXT, KEK, and JAXA (Japan), and the K.A.~Wallenberg Foundation, the Swedish Research Council and the National Space Board (Sweden). Science analysis support in the operations phase from INAF (Italy) and CNES (France) is also gratefully acknowledged. This work performed in part under DOE Contract DE-AC02-76SF00515.

Additional support for science analysis during the operations phase is gratefully acknowledged from the Istituto Nazionale di Astrofisica in Italy and the Centre National d'Etudes Spatiales in France.

MHG, MLG and SG acknowledge support from Agence Nationale de la Recherche (grant ANR- 17-CE31-0014).

The contribution of JDG is supported by the National Aeronautics and Space Administration (NASA) under grant number NNX17AL74G issued through the NNH16ZDA001N Astrophysics Data Analysis Program (ADAP). JDG is also supported by the NYU Abu Dhabi Research Enhancement Fund (REF) under grant RE022 and  NYU Abu Dhabi Grant AD022.

SG acknowledges support from the Observatory of Paris (Action F\'ed\'eratrice CTA).

We thank David A. Smith for useful discussions.

C.J.C. acknowledges support from the ERC under the European Union’s Horizon 2020 research and innovation programme (grant agreement No. 715051; Spiders).

This research has made use of NASA's Astrophysics Data System Bibliographic Services.

\appendix
\section{Pulsar Wind Nebula Modeling of HESS J1640\texorpdfstring{$-$}{-}465}
\label{sec:pwndetails}

As mentioned in Section \ref{sec:pwn}, the best way of testing a PWN origin for the $\gamma$-rays observed from HESS J1640$-$465 is to determine if the dynamical and spectral properties of this source can be reproduced by a model for the evolution of a PWN inside a SNR.  We use a model similar to that described by \citet{gelfand09}, which successfully reproduced the properties of other $\gamma$-ray emitting PWNe such as G54.1+0.3 \citep{gelfand15} and Kes 75 \citep{gelfand14}.  In this ``one-zone'' model, the magnetic field strength and particle distribution inside the PWN are assumed to be uniform (e.g., \citealt{reynolds84, gelfand09}), and the SNR is expanding inside a medium with uniform number density $n_{\rm ism}$.  Furthermore, it assumes the density profile of the unshocked SN ejecta consists of a uniform density $(\rho \propto r^0)$ core surrounded by a $\rho \propto r^{-9}$ envelope (e.g., \citealt{blondin01, gelfand09}), where $r$ is the distance from the explosion site (assumed to be the center of the SNR).  While different supernova progenitors will likely result in ejecta with different density profiles, this does not substantially affect the evolution of the PWN \citep{chevalier05}. In some cases, it is possible to assume the dependence between the ISM density $n_{\rm ism}$, initial kinetic energy $E_{\rm sn}$ and mass of the ejecta $M_{\rm ej}$ as predicted by self-similar evolution of the SNR.  However, this is not possible for HESS J1640$-$465 for two reasons:
\begin{enumerate}
    \item We are attempting to reproduce the size of both the PWN and the SNR.  Before the PWN collides with the SNR's reverse shock, its size is sensitive to the density of the innermost ejecta, whose dependence of $E_{\rm sn}$ and $M_{\rm ej}$ differs from the dependence of the SNR's forward shock on these parameters (and $n_{\rm ism}$).  Therefore, two different parameterizations are needed, which increases the number of model parameters.
    \item The large size of the PWN relative to the SNR shell in this system ($\theta_{\rm pwn} \sim 0.75 \theta_{\rm snr}$; Table \ref{tab:pwn_obsprop}) suggests that it might have already collided with the SNR reverse shock.  In this case, the evolution of the PWN is sensitive to a number of different parameters (e.g., the time of the collision and the pressure of the reverse-shocked material; see Gelfand et al.\ 2009).  While our model suggests this collision has not happened quite yet, there was no basis for rejecting this possibility a priori.
\end{enumerate}

In this model, the pulsar injects energy into the PWN at a rate, $\dot{E}$, which evolves as (e.g., \citealt{goldreich69, pacini73}):
\begin{eqnarray}
\label{eqn:edot}
\dot{E}(t) & = & \dot{E}_0 \left(1 + \frac{t}{\tau_{\rm sd}}  \right)^{-\frac{p+1}{p-1}}
\end{eqnarray}
where $\dot{E}_0$ is the initial spin-down luminosity, p is the braking index, and $\tau_{\rm sd}$ is the spin-down timescale (equivalent to the ``characteristic age'' at birth) of the associated pulsar PSR J1640$-$4631.  As listed in Table \ref{tab:pwn_obsprop}, timing observations of this neutron star have determined its current braking index, characteristic age $t_{\rm ch}$, and spin-down luminosity.  Therefore, for a particular value of $\tau_{\rm sd}$, we can determine the ``true'' age $t_{\rm age}$:
\begin{eqnarray}
\label{eqn:tage}
t_{\rm age} & = & \frac{2 t_{\rm ch}}{p-1}-\tau_{\rm sd} 
\end{eqnarray}
of this system as well as $\dot{E}_0$:
\begin{eqnarray}
\label{eqn:e0dot}
\dot{E}_0 & = & \dot{E} \left(1+\frac{t_{\rm age}}{\tau_{\rm sd}} \right)^{+\frac{p+1}{p-1}}
\end{eqnarray}
assuming that $p$ is constant with time.  This allows us to decrease the number of model parameters by two.

Furthermore, this model assumes that a constant fraction $\eta_{\rm B}$ of this energy is injected into the PWN in the form of magnetic fields, while the remainder ($1-\eta_{\rm B}$) is injected as relativistic e$^{\pm}$ whose spectrum is described by a broken power-law 
\begin{eqnarray}
   \label{eqn:bpl}
    \frac{{\rm d}\dot{N}}{{\rm d}E} & = & \left\{ \begin{array}{cc}
    \dot{N}_{\rm break} \left(\frac{E}{E_{\rm break}} \right)^{-p_1} & E_{\rm min} < E < E_{\rm break} \\
    \dot{N}_{\rm break} \left(\frac{E}{E_{\rm break}} \right)^{-p_2} & E_{\rm break} < E < E_{\rm max} \\
    \end{array} \right.
 \end{eqnarray}
where the five free parameters ($E_{\rm min}$, $E_{\rm break}$, $E_{\rm max}$, $p_1$, and $p_2$) in Equation \ref{eqn:bpl} are assumed to be constant with time and the normalization $\dot{N}_{\rm break}$ is calculated by requiring that:
\begin{eqnarray}
   (1-\eta_{\rm B})\dot{E} & = & \int\limits_{E_{\rm min}}^{E_{\rm max}} E \frac{{\rm d}\dot{N}}{{\rm d}E} {\rm d}E
\end{eqnarray}
at all times $t$. As the PWN evolves, the energy of these particles changes due to adiabatic expansion / contraction of the nebulae (e.g., \citealt{gelfand09}):
\begin{eqnarray}
\dot{E}_{\rm ad}(t,E) & = & -\frac{\dot{R}_{\rm pwn}(t)}{R_{\rm pwn}(t)} E
\end{eqnarray}
where $R_{\rm pwn}$ is the radius of the (assumed spherical) PWN, calculated using the prescription described by \citet{gelfand09}, and radiative losses resulting from the synchrotron and inverse Compton emission of this particles.  The synchrotron emission is calculated assuming that, inside the PWN, the angle between a particle's velocity and local magnetic field is randomly distributed, and the evolution of the PWN's magnetic field is calculated assuming magnetic flux conservation as described by \citet{gelfand09}.  For inverse Compton emission, we assume the particles are scattering photons from the Cosmic Microwave Background (temperature $T_{\rm cmb} = 2.7255~{\rm K}$; \citealt{fixsen09}) as well as additional background fields with a blackbody spectrum and normalization $K_{\rm ic}$, such that this photon field has an energy density:
\begin{eqnarray}
u_{\rm ic} & = & K_{\rm ic} a_{\rm bb} T_{\rm ic}^4,
\end{eqnarray}
where $a_{\rm bb} \approx 7.5657\times10^{-15}~$erg ${\rm cm^{-3} K^{-4}}$.

\begin{table}[ht]
    \centering
    \begin{tabular}{cccc}
    \hline
    \hline
    {\sc Property} & {\sc Observed Value} & {\sc Model Prediction} & {\sc Citation} \\
    \multicolumn{4}{c}{\it PSR J1640$-$4631} \\
    Current Spin-down Luminosity $\dot{E}$ & $4.4\times10^{36}~$erg ${\rm s^{-1}}$ & $\cdots$ & \cite{gotthelf_nustar_2014} \\  
    Characteristic Age $t_{\rm ch}$ &  3350~years & $\cdots$ & \cite{gotthelf_nustar_2014} \\
    Braking Index $p$ & $3.15\pm0.03$ & $\equiv3.15$ & \cite{archibald16} \\
    \hline
    \multicolumn{4}{c}{\it SNR G338.3$-$0.0} \\
    Radius $\theta_{\rm snr}$ & $4.\!^{\prime}45 \pm 0.\!^{\prime}5$ & $4.\!^{\prime}2$ & \cite{shaver70} \\
    \hline
    \multicolumn{4}{c}{\it PWN J1640$-$465} \\
    Radius $\theta_{\rm pwn}$ & $3.\!^{\prime}3 \pm 0.\!^{\prime}2$ & $3.\!^{\prime}5$ & \cite{lemiere_high-resolution_2009} \\
    660~MHz Flux Density $S_660$ & $<690\pm300$~mJy & 70~mJy & \cite{castelletti_first_2011} \\
    $2-25$~keV Unabsorbed Flux & $(1.7\pm0.4)\times10^{-12}~$(erg ${\rm cm^{-2}~s^{-1}})^{*}$ & $1.67\times10^{-12}~$(erg ${\rm cm^{-2}~s^{-1}})^{*}$ & \cite{gotthelf_nustar_2014} \\
    $2-25$~keV Photon Index & $2.2_{-0.4}^{+0.7}$ & $2.58$ & \cite{gotthelf_nustar_2014} \\
    $\nu F_{\nu}({\rm 0.35~GeV})$ & $(1.32 \pm1.57)\times10^{-12}~*$ & $0.76\times10^{-13}~*$ & $\cdots$ \\
    $\nu F_{\nu}({\rm 0.785~GeV})$ & $1.38_{-2.24}^{+2.19}\times10^{-12}~*$ & $1.27\times10^{-12}~*$ & $\cdots$ \\
    $\nu F_{\nu}({\rm 1.68~GeV})$ & $3.22_{-1.48}^{+1.46}\times10^{-12}~*$ & $1.99\times10^{-12}~*$ & $\cdots$ \\
    $\nu F_{\nu}({\rm 3.59~GeV})$ & $3.64_{-1.02}^{+1.02}\times10^{-12}~*$ & $2.96\times10^{-12}~*$ & $\cdots$ \\
    $\nu F_{\nu}({\rm 7.67~GeV})$ & $3.01_{-1.02}^{+1.02}\times10^{-12}~*$ & $4.15\times10^{-12}~*$ & $\cdots$ \\
    $\nu F_{\nu}({\rm 16.4~GeV})$ & $3.54_{-1.03}^{+1.03}\times10^{-12}~*$ & $5.49\times10^{-12}~*$ & $\cdots$ \\
    $\nu F_{\nu}({\rm 35~GeV})$ & $6.31_{-1.61}^{+1.61}\times10^{-12}~*$ & $6.77\times10^{-12}~*$ & $\cdots$ \\
    $\nu F_{\nu}({\rm 75~GeV})$ & $8.81_{-2.75}^{+2.75}\times10^{-12}~*$ & $6.98\times10^{-12}~*$ & $\cdots$ \\
    $\nu F_{\nu}({\rm 160~GeV})$ & $5.65_{-2.71}^{+2.71}\times10^{-12}~*$ & $7.15\times10^{-12}~*$ & $\cdots$ \\
    $\nu F_{\nu}({\rm 289~GeV})$ & $5.34_{-2.42}^{+2.27}\times10^{-12}~*$ & $7.05\times10^{-12}~*$ & \cite{abramowski_hess_2014} \\
    $\nu F_{\nu}({\rm 340~GeV})$ & $<6.47\times10^{-12}~*$ & $6.38\times10^{-12}~*$ & $\cdots$ \\
    $\nu F_{\nu}({\rm 350~GeV})$ & $5.92_{-1.04}^{+1.01}\times10^{-12}~*$ & $6.50\times10^{-12}~*$ & \cite{abramowski_hess_2014} \\
    $\nu F_{\nu}({\rm 424~GeV})$ & $6.99_{-0.80}^{+0.78}\times10^{-12}~*$ & $6.10\times10^{-12}~*$ & \cite{abramowski_hess_2014} \\
    $\nu F_{\nu}({\rm 513~GeV})$ & $5.57_{-0.62}^{+0.60}\times10^{-12}~*$ & $5.52\times10^{-12}~*$ & \cite{abramowski_hess_2014} \\
    $\nu F_{\nu}({\rm 622~GeV})$ & $5.66_{-0.59}^{+0.57}\times10^{-12}~*$ & $5.95\times10^{-12}~*$ & \cite{abramowski_hess_2014} \\
    $\nu F_{\nu}({\rm 734~GeV})$ & $5.08_{-6.36}^{+6.36}\times10^{-12}~*$ & $5.20\times10^{-12}~*$ & $\cdots$ \\
    $\nu F_{\nu}({\rm 753~GeV})$ & $6.37_{-0.57}^{+0.55}\times10^{-12}~*$ & $5.34\times10^{-12}~*$ & \cite{abramowski_hess_2014} \\
    $\nu F_{\nu}({\rm 913~GeV})$ & $5.60_{-0.52}^{+0.50}\times10^{-12}~*$ & $5.04\times10^{-12}~*$ & \cite{abramowski_hess_2014} \\
    $\nu F_{\nu}({\rm 1.11~TeV})$ & $5.22_{-0.50}^{+0.49}\times10^{-12}~*$ & $4.38\times10^{-12}~*$ & \cite{abramowski_hess_2014} \\
    $\nu F_{\nu}({\rm 1.34~TeV})$ & $5.57_{-0.51}^{+0.49}\times10^{-12}~*$ & $4.20\times10^{-12}~*$ & \cite{abramowski_hess_2014} \\
    $\nu F_{\nu}({\rm 1.62~TeV})$ & $4.59_{-0.49}^{+0.47}\times10^{-12}~*$ & $4.28\times10^{-12}~*$ & \cite{abramowski_hess_2014} \\
    $\nu F_{\nu}({\rm 1.97~TeV})$ & $3.99_{-0.48}^{+0.46}\times10^{-12}~*$ & $3.91\times10^{-12}~*$ & \cite{abramowski_hess_2014} \\
    $\nu F_{\nu}({\rm 2.38~TeV})$ & $2.87_{-0.45}^{+0.43}\times10^{-12}~*$ & $3.44\times10^{-12}~*$ & \cite{abramowski_hess_2014} \\
    $\nu F_{\nu}({\rm 2.89~TeV})$ & $3.77_{-0.50}^{+0.48}\times10^{-12}~*$ & $3.20\times10^{-12}~*$ & \cite{abramowski_hess_2014} \\
    $\nu F_{\nu}({\rm 3.50~GeV})$ & $3.24_{-0.50}^{+0.48}\times10^{-12}~*$ & $2.73\times10^{-12}~*$ & \cite{abramowski_hess_2014} \\
    $\nu F_{\nu}({\rm 4.24~TeV})$ & $2.65_{-0.51}^{+0.48}\times10^{-12}~*$ & $2.56\times10^{-12}~*$ & \cite{abramowski_hess_2014} \\
    $\nu F_{\nu}({\rm 5.13~TeV})$ & $1.96_{-0.48}^{+0.45}\times10^{-12}~*$ & $2.63\times10^{-12}~*$ & \cite{abramowski_hess_2014} \\
    $\nu F_{\nu}({\rm 6.22~TeV})$ & $1.70_{-0.49}^{+0.45}\times10^{-12}~*$ & $2.14\times10^{-12}~*$ & \cite{abramowski_hess_2014} \\
    $\nu F_{\nu}({\rm 7.53~TeV})$ & $1.95_{-0.55}^{+0.50}\times10^{-12}~*$ & $2.04\times10^{-12}~*$ & \cite{abramowski_hess_2014} \\
    $\nu F_{\nu}({\rm 10.2~TeV})$ & $8.77_{-3.53}^{+3.21}\times10^{-13}~*$ & $1.55\times10^{-12}~*$ & \cite{abramowski_hess_2014} \\
    \hline
    \hline
    \end{tabular}
    \caption{Observed properties of HESS J1640$-$465, as well as value predicted by our model for the combination of input parameters given in Table \ref{tab:pwn_model_pars}.  The errors for the values of $\nu F_\nu$ in the $\gamma$-ray regime are the square-root of the sum of the squares of the statistical and systematic errors. The $\cdots$ indicate quantities whose values were fixed during the modeling and the * indicate that the units are in erg ${\rm cm^{-2}~s^{-1}}$.}
    \label{tab:pwn_obsprop}
\end{table}

Similar to the process described by \citet{gelfand15}, we used a Metropolis Monte Carlo Markoff Chain  \citep[e.g.][]{metropolis85} to determine the combination of the 15 model input parameters $\Theta$, listed in Table \ref{tab:pwn_model_pars}, the best reproduces the 35 observed properties $\mathcal{D}$ listed in Table \ref{tab:pwn_obsprop}.  This was done by identifying the combination $\theta$ whose predicted observable quantities $\mathcal{M}$ resulted in the lowest $\chi^2$, defined as:
\begin{eqnarray}
\label{eqn:chi2}
\chi^2 & = & \sum\limits_{i=1}^{i=35} \left(\frac{{\mathcal D}_i - \mathcal{M}_i}{\sigma_i} \right)^2,
\end{eqnarray}
where $\sigma_i$ is the error on observed quantity ${\mathcal D}_i$ listed in Table \ref{tab:pwn_obsprop}. While such a procedure can be used to identify degeneracies between the model parameters as well as estimate their errors, the necessary mapping of the possible parameter space  requires $\gtrsim10^6$ trials (e.g. Gelfand et al. 2015), which is beyond the scope of this work.


\bibliographystyle{apj}
%\bibliography{biiblio}
\bibliography{sample63.bib}


\end{document}